\documentclass{article}

\usepackage{arxiv}

\usepackage[utf8]{inputenc} 
\usepackage[T1]{fontenc}  
\usepackage{hyperref}    
\usepackage{url}      
\usepackage{booktabs}    
\usepackage{amsfonts}    
\usepackage{nicefrac}    
\usepackage{microtype}   
\usepackage{graphicx}
\usepackage{natbib}
\usepackage{doi}

\usepackage{algorithm}
\usepackage[noend]{algpseudocode}
\usepackage{amsmath}
\usepackage{hyperref}

\usepackage[table,xcdraw]{xcolor}

\usepackage{caption}

\title{fdrSAFE: Selective Aggregation for \\Local False Discovery Rate Estimation}

\date{} 					

\author{ 
  \href{https://orcid.org/0000-0002-1411-8750}{\includegraphics[scale=0.06]{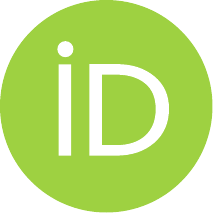}\hspace{1mm}Jenna M. Landy} \\
	Harvard University\\
	\texttt{jlandy@g.harvard.edu} \\
	\And
  \href{https://orcid.org/0000-0002-8783-5961}{\includegraphics[scale=0.06]{orcid.pdf}\hspace{1mm}Giovanni Parmigiani} \\
	Dana Farber Cancer Institute\\
    Harvard University\\
    \texttt{gp@jimmy.harvard.edu} \\
}

\renewcommand{\shorttitle}{fdrSAFE: Selective Aggregation for fdr Estimation}

\hypersetup{
pdftitle={fdrSAFE: Selective Aggregation for fdr Estimation},
pdfsubject={stat-methods},
pdfauthor={Jenna Landy, Giovanni Parmigiani},
pdfkeywords={False Discovery Rates, Multiple Hypothesis Testing, Model Selection, Ensembling},
}

\begin{document}
\maketitle

\begin{abstract}
Estimating local false discovery rates (fdr) is central to large-scale multiple hypothesis testing, yet different methods often produce divergent results, and there is little guidance for selecting among them. Because ground truth hypothesis labels are unobservable, standard model selection cannot be used. We present fdrSAFE (selective aggregation for fdr estimation), a data-driven selective ensembling approach that estimates model performances on synthetic datasets designed to resemble the observed data but with known ground truth. With simulation studies and an experimental spike-in transcriptomic dataset, we show that fdrSAFE achieves robust near-optimality, performing well across diverse settings where baseline model performances vary. Along with improved fdr estimates, this framework enhances replicability by replacing arbitrary model choice with a principled, data-adaptive procedure. An open-source R software package is available on GitHub at \href{https://github.com/jennalandy/fdrSAFE}{jennalandy/fdrSAFE}.

\end{abstract}


\section{Introduction}
Multiple hypothesis testing is common when working with high-dimensional data across a range of scientific fields. For example, in high-throughput transcriptomics, a common question is which of many genes are differentially expressed between two groups, such as two cell types or individuals with and without a disease. When testing many hypotheses at once---typically thousands---the number of falsely rejected hypotheses can grow unreasonably large, and accounting for multiplicities is essential for reliable inference \citep{Dudoit2003, farcomeni2008review}. A commonly used approach in this setting is the estimation of local false discovery rates (fdr) \citep{efron2001}, which quantify the probability that a hypothesis is null conditional on its test statistic.

Modeling assumptions play a central role in fdr estimation, and different choices can yield very different fdr estimates. Across popular R software packages {\tt locfdr} \citep{locfdr}, {\tt fdrtool} \citep{fdrtool, strimmer}, and {\tt qvalue} \citep{qvalue}, our simulation studies show that no single model performs best---different approaches excel depending on the dataset. Package choice is consequential, yet often arbitrary as there is limited practical guidance on which package to use and under what conditions. Choosing among fdr estimation methods is a model selection problem, but is complicated by the unsupervised nature of hypotheses, which means a model's performance cannot be estimated empirically. In this paper, we address this key challenge of evaluating and selecting fdr estimation models in the absence of ground-truth hypothesis labels.

The remainder of the paper is organized as follows. After reviewing relevant literature on false discovery rate control (Sections \ref{sec:fdr}, \ref{sec:est}), we formalize this challenge as a model selection problem (Section \ref{sec:modsel}). We then introduce fdrSAFE (selective aggregation for fdr estimation, Section \ref{sec:methods}), an approach to estimate fdr that evaluates model performance on synthetic data and performs a selective ensemble, or aggregation, of well-performing models. fdrSAFE consists of three components: (1) generating synthetic datasets that mimic observed data but with known ground truth, (2) evaluating model performance on these datasets and selecting a high-performing subset of models, and (3) computing a weighted ensemble of the selected models’ estimates on the observed data. Through simulation studies (Section \ref{sec:simulations}) and an experimental transcriptomics application (Section \ref{sec:experiment}), we show that fdrSAFE outperforms three popular R software packages with default settings. These results also serve as a rigorous comparison of these baseline fdr packages under default settings, highlighting their relative strengths and limitations in practical applications. Further, by avoiding arbitrary model choice, fdrSAFE improves replicability in large-scale multiple testing. An open-source R software package implementing fdrSAFE is available on GitHub at \href{https://github.com/jennalandy/fdrSAFE}{jennalandy/fdrSAFE}.

\section{Definitions}

\subsection{False Discovery Rate Methodologies}\label{sec:fdr}

Early multiple test correction approaches control the Type-I error rate for a set of hypotheses, for example adjusting p-values or significance cutoffs to attain a pre-specified family-wise error rate (FWER) \citep{tukey1953problem,Tukey:1991ga} or a global false discovery rate (FDR) \citep{benj.hoch}. Alternatively, tail-end (Fdr) \citep{efro:tibs:2002} and local (fdr) \citep{efron2001} false discovery rates can be reported separately for each hypothesis and can be used in place of p-values to quantify evidence against the null. All of these quantities are unknown so must be estimated or bounded for practical application. Following \citet{efron2001} and \citet{efron2007}, we use the standard notation of alternate capitalization for local (fdr), tail-end (Fdr), and global (FDR) false discovery rates.

Formally, consider a collection of $I$ null hypotheses $\mathbf H_0 = (H_{01}, \ldots, H_{0I})$, associated test statistics $\mathbf u = (u_1, \dots, u_I)$, and rejection regions $\mathbf R = (R_1, \dots, R_I)$ produced by some testing procedure. The global false discovery rate (FDR) is the unknown expected proportion of null hypotheses among those whose test statistics fall into their rejection regions \citep{benj.hoch}, that is:
  \begin{align}
    \mbox{global false discovery rate: } \mbox{FDR}&= Pr(H_0 | U \in R)\\
    & \approx
    \frac{ \sum_{i = 1}^I \mathbf{1} \{ u_i \in R_i \cap H_{0i} \} }
    {\sum_{i=1}^I \mathbf{1} \{ u_i \in R_i \} } \label{eq:FDR},
  \end{align}
where $\mathbf{1}$ is a binary indicator of its argument.

The tail-end false discovery rate (Fdr) for hypothesis $i$ is the unknown expected proportion of hypotheses that are null among those with test statistics as or more extreme than $u_i$ \citep{efro:tibs:2002}. When large magnitudes of $u$ denote more extreme departures from the null, we have:
  \begin{align}
    \mbox{tail-end false discovery rate: Fdr}_i &= Pr\left(H_{0} \big| | U| \ge |u_i|\right)\\
    &\approx\frac{\sum_{i' = 1}^I \mathbf{1} \{ |u_{i'}| \ge |u_i| \cap H_{0i'} \} }
    {\sum_{i' = 1}^I \mathbf{1} \{ |u_{i'}| \ge |u_i| \} } \label{eq:Fdr}
  \end{align}
Quantities (\ref{eq:FDR}) and (\ref{eq:Fdr}) are defined as $0$ when their denominator is $0$.

The local false discovery rate (fdr) for hypothesis $i$ is the unknown probability of a hypothesis being null given a test statistic $u_i$ \citep{efron2001}. A commonly made assumption is that the test statistics come from a continuous mixture distribution with components $f_0(u)$ and $f_1(u)$, describing the distributions of test statistics generated by null and alternative hypotheses, respectively. The mixing parameter $\pi_0$ is the unknown proportion of null hypotheses. Then, local false discovery rates can be defined with Bayes rule as: 
  \begin{equation}
  \label{eq:fdr_def}
    \begin{aligned}
    \mbox{local false discovery rate: fdr}_i &= Pr(H_0|U = u_i) = \frac{\pi_0f_0(u_i)}{f(u_i)},
    \end{aligned}
  \end{equation}
where $f(u_i) = \pi_0f_0(u_i) + (1-\pi_0)f_1(u_i)$.

\cite{efro:tibs:2002} also show a simple relation between tail-end and local false discovery rates:
    \begin{align}
        \mbox{Fdr}_i& = E\left[ \hbox{fdr}_{i'}\Big| |u_{i'}| \ge |u_i| \right].\label{eq:Fdr_relation}
    \end{align}

\subsection{Estimating fdr}\label{sec:est}
\label{sec:estimatingfdr}
In most implementations, local false discovery rates (fdrs) are estimated from the test statistics vector $\mathbf{u}$ alone \citep{locfdr, fdrtool, qvalue}, though important variations exist \citep{Stephens:2017en}. Estimating fdrs takes advantage of the large number of hypotheses with an empirical Bayes approach \citep{efron2001}, first estimating the mixture $f$ and the weighted null $\pi_0 f_0$ using all statistics $\mathbf{u}$, then evaluating Equation~\ref{eq:fdr_def} at each $u_i$ to obtain individual fdr estimates.

Estimated $\pi_0$ and fdr values depend on all test statistics $\mathbf u$. We denote this dependence by defining estimates as functions of $\mathbf u$. In all equations, we use boldface to denote vectors and the subscript $\theta$ to refer to a particular model:
\begin{align}
    \widehat{\mbox{fdr}}_{\theta, i}(\mathbf u) &= \frac{\hat\pi_{0, \theta}(\mathbf u)\hat f_{0, \theta}(u_i|\mathbf u)}{\hat f_\theta(u_i|\mathbf u)}\\
    \widehat{\mbox{\textbf{fdr}}}_{\theta}(\mathbf u) &= \left(\widehat{\mbox{fdr}}_{\theta, 1}(\mathbf u),..., \widehat{\mbox{fdr}}_{\theta, I}(\mathbf u)\right).
\end{align}

Estimates of local fdr can be leveraged to additionally estimate tail-end Fdr and global FDR. Using Equation \ref{eq:Fdr_relation}, estimates of local fdr can easily be converted into estimates of tail-end Fdr via:
\begin{equation}
\begin{aligned}
   \label{eq:Fdr_fdr_rel} 
  \widehat{\mbox{Fdr}}_{\theta, i}(\mathbf u)&=
      \frac{\sum_{i' = 1}^I 
      \mathbf{1} \{ |u_{i'}| \ge |u_i| \} \widehat{\hbox{fdr}}_{\theta,i'}(\mathbf u) }
    {\sum_{i' = 1}^I \mathbf{1} \{ |u_{i'}| \ge |u_i| \} }\\
    \widehat{\mbox{\textbf{Fdr}}}_{\theta}(\mathbf u) &= \left(\widehat{\mbox{Fdr}}_{\theta, 1}(\mathbf u), ..., \widehat{\mbox{Fdr}}_{\theta, I}(\mathbf u)\right).
\end{aligned}
\end{equation}
Additionally, if the rejection region is defined by a cutoff point $u_j$, such that $R_i = \{ |u_i| \ge |u_j| \}$ for all tests $i$, then $\widehat{\mbox{Fdr}}_{\theta, j} (\mathbf u)$ estimates global FDR. These quantities can also be estimated using Bayesian decision theoretic approaches~\citep{muel:parm:rice:2007}.

Different R software packages, and within them specifications of parameter values, use different estimation methods, distributional assumptions, and subsets of the data in estimating $f$ and $\pi_0 f_0$. Importantly, different models can return very different estimates, and thus can lead to different conclusions. 

\subsection{Model Selection and Ensembling}\label{sec:modsel}

Model selection is used to identify one or a subset of models which generalize well to $\mathcal{P}$, the true data generating distribution. The goal is to minimize the objective, or the expectation of some loss $L$, $\mathcal{L} = E_{\mathcal{P}}[L(\widehat{\mbox{\textbf{fdr}}}_\theta(\mathbf u), \mbox{\textbf{fdr}})]$.

A straightforward approach in model selection is a grid search, which exhaustively considers all combinations of selected parameter values and orders models from best to worst by estimated objective \citep{yu2020hyper}. When true outputs can be observed, the objective can be estimated empirically by the loss in a held-out portion of the data \citep{kuhn2013applied}. In some cases, the objective is estimated by the loss in external datasets \citep{nomura2021} or synthetic datasets constructed by bootstrap or jackknife resampling \citep{efron1982jackknife, kohavi1995study}. However, true fdr is unknown, making this challenging.

An alternative to selecting a single model is to combine, or ensemble, estimates from multiple models. The literature on ensembling is extensive \citep{Dong2020} and shows that ensembling over all possible models with weights corresponding to model performance can improve predictive ability over using a single model. Further improvement can be seen when being selective to only include well-performing models in an ensemble, as, for example, in Occam's window~\citep{MadiganRaftery1994}.

\section{Methods}
\label{sec:methods}

\subsection{fdrSAFE: Selective Aggregation for fdr Estimation}

The motivation for this work is the need for a practical and data-driven approach to estimate fdr in the current landscape. Models disagree, and their relative merits vary widely across data applications. Because fdr is not observed, model performances cannot be assessed empirically. Our proposed solution, fdrSAFE, is a selective ensembling algorithm for estimating fdr in large-scale multiple hypothesis testing. 

fdrSAFE has three main steps. First, we generate synthetic datasets with test statistics $\mathbf{u}$ and ground truth $\mbox{\textbf{fdr}}$ through a parametric bootstrap, where the parametric synthetic generator is fit on observed test statistics. Next, we estimate model performances using these datasets and select a high performing subset of models. Finally, on the observed data, we ensemble the estimates from selected models with weights proportional to estimated model performances. Algorithm \ref{alg:framework} outlines the method, and the following sections include details for each of the three steps.

Our work focuses on two-sided hypotheses where the distribution of test statistics is centered at zero under the null and large magnitudes provide evidence against the null. In the context of differential gene expression, this includes two-sample t-test statistics \citep{ryu2002relationships, cui2003statistical} and log fold change estimates from over-dispersed log-linear models \citep{lu2005identifying, robinson2007moderated} such as from R software packages {\tt edgeR} \citep{chen2020edger} and {\tt DESeq2} \citep{varet2016sartools}.

We consider fdr models implemented by two R packages that estimate fdr from this type of test statistic: {\tt locfdr} \citep{locfdr} and {\tt fdrtool} \citep{fdrtool, strimmer}. A third R package, {\tt qvalue} \citep{qvalue}, instead takes in p-values, but we still recognize it as a widely used and valuable implementation. We incorporate qvalue by converting statistics $\mathbf{u}$ to p-values. In the fdrSAFE R package, the user may provide their own function to convert statistics to p-values, with the default as a t-distributed null (or a standard Normal if degrees of freedom are not provided).

We define a grid of models which includes the three packages with all possible combinations of categorical parameters and equally spaced choices of numeric parameters, resulting in $M=292$ models. For a given dataset, we exclude models that do not converge or produce errors.

\begin{algorithm}
\caption{-- \textbf{fdrSAFE}\\
\textbf{Input}: Observed test statistics vector $\mathbf{u}$, number of synthetic datasets $N$, ensemble size $m$, grid of model specifications $\Theta$ where $||\Theta|| = M \geq m$\\
\textbf{Output}: $\widehat{\textbf{fdr}}_{\mbox{{\scriptsize SAFE}}}(\mathbf u)$, $\widehat{\textbf{Fdr}}_{\mbox{{\scriptsize SAFE}}}(\mathbf u)$ and $\widehat{\pi}_{0,\mbox{{\scriptsize SAFE}}}(\mathbf u)$}\label{alg:framework}

\begin{algorithmic}[1]
\Statex \hspace{-5mm} \textbf{\# Step 1: Synthetic Generator}
\State $\hat{\phi} \gets$ MLE of $\phi$ under the synthetic generator
with data $\mathbf{u}$ via Expectation-Maximization (EM) algorithm
\For{$n = 1, \dots, N$}
\State Draw synthetic data $\{\mathbf{u}^n, \mbox{\textbf{fdr}}^n, \mathbf{l}^n\}$ given synthetic generator $\hat{\phi}$
\EndFor
\Statex 
\Statex \hspace{-4mm}\textbf{\# Step 2: Model Subset Selection}
\For{fdr model $\theta \in \Theta$}
\For{synthetic dataset $n \in 1,...,N$}
\State Fit fdr model $\theta$ to synthetic dataset $\{\mathbf{u}^n, \mbox{\textbf{fdr}}^n, \mathbf{l}^n\}$ to obtain $\widehat{\mbox{\textbf{fdr}}}_\theta(\mathbf{u}^n)$ and $\hat \pi_{0, \theta}(\mathbf{u}^n)$
\EndFor
\State Estimate objective $\widehat{\mathcal{L}}_{\text{SAFE}}(\theta) \gets \frac{1}{N}\sum_{n = 1}^N \sum_{i = 1}^I(\widehat{\mbox{fdr}}_{\theta,i}(\mathbf{u}^n) - \mbox{fdr}_i^n )^2$
\EndFor
\State 
Sort fdr models such that 
$\widehat{\mathcal{L}}_{\text{SAFE}}(\theta_{(1)}) \le \widehat{\mathcal{L}}_{\text{SAFE}}(\theta_{(2)}) \le \dots \le \widehat{\mathcal{L}}_{\text{SAFE}}(\theta_{(M)})$ 
\State Select $m$ top performing models $\Theta^* = \{\theta_{(1)},\dots,\theta_{(m)}\}$
\Statex 
\Statex \hspace{-4mm}\textbf{\# Step 3: Ensemble}
\State Compute ensemble weights $w_{\theta} = \frac{1-\widehat{\mathcal{L}}_{\text{SAFE}}(\theta)}{\sum_{\theta' \in \Theta^*} \left(1 -\widehat{\mathcal{L}}_{\text{SAFE}}(\theta')\right)}$ \hspace{2mm} $\forall \theta \in \Theta^*$
\State $\widehat\pi_{0, \mbox{{\scriptsize SAFE}}}(\mathbf u) = \sum_{\theta \in \Theta^*} w_{\theta}\widehat{\pi}_{0,\theta}(\mathbf{u})$
\State $\widehat{\mbox{\textbf{fdr}}}_{\mbox{{\scriptsize SAFE}}}(\mathbf u) = \sum_{\theta \in \Theta^*} w_{\theta} \; \widehat{\mbox{\textbf{fdr}}}_\theta(\mathbf{u})$
\State $\widehat{\hbox{Fdr}}_{i, \mbox{{\scriptsize SAFE}}}(\mathbf u) =  \frac{\sum_{i' = 1}^I 
      \mathbf{1} \{ |u_{i'}| \ge |u_i| \} \widehat{\hbox{fdr}}_{i'}(\mathbf u) }
    {\sum_{i' = 1}^I \mathbf{1} \{ |u_{i'}| \ge |u_i| \} },$ \quad $\forall i \in 1, \dots, I$
\Statex 
\State Return $\widehat{\text{\textbf{fdr}}}_{\mbox{{\scriptsize SAFE}}}$, $\widehat{\textbf{Fdr}}_{\mbox{{\scriptsize SAFE}}}$, 
and $\widehat{\pi}_{0,\mbox{{\scriptsize SAFE}}}$
\end{algorithmic}
\end{algorithm}

\subsection{Step 1: Synthetic Data Generation}

The \textbf{synthetic generator} is used to generate synthetic datasets that resemble the observed data in key ways but with known ground truth. We do not recommend it as an fdr estimation model. In fact, fdr estimates computed directly from the synthetic generator are often inferior to those of fdrSAFE and individual component methods. Instead, its role is to enable performance evaluation of fdr models by creating synthetic data where ground truth is available. If a synthetic generator distribution is sufficiently similar to the real data distribution, then models performing well on synthetic data are likely to perform well on the observed data too.

We define the synthetic generator, referred to by subscript $G$, as a mixture of two components. The first, $f_{G0}(u|\sigma_0)$, corresponding to statistics under the null hypothesis, is a Normal distribution centered at zero with standard deviation $\sigma_0$ (Equation \ref{eq:wm_null}). The second, $f_{G1}(u|\pi_{1n},\sigma_{1n}, \sigma_{1p})$, corresponding to statistics under the alternative hypothesis, is a mixture of two nonlocal Half-Normal components, inspired by nonlocal priors from \citet{Rossell2017}, with opposite signs and separate standard deviations for the negative and positive components, $\sigma_{1n}$ and $\sigma_{1p}$. Mixing parameter $\pi_{1n}$ controls the proportion of alternative test statistics that are negative (Equation \ref{eq:wm_alt}). This allows for asymmetry in not-null test statistics, both in terms of mass and spread, while keeping density away from zero. The full synthetic generator mixture is completed by the mixture weight $\pi_0$, the proportion of null tests. For conciseness, $\phi = (\pi_0, \sigma_0,\pi_{1n},\sigma_{1n}, \sigma_{1p})$, and $f_{G}(u|\phi)$ is the implied marginal (Equation \ref{eq:wm}).
\begin{align}
  &f_{G,0}(u|\sigma_0) = N(u; 0, \sigma_0)\label{eq:wm_null}\\
  &f_{G,1}(u|\pi_{1n},\sigma_{1n}, \sigma_{1p}) = \pi_{1n} \cdot \frac{2u^2}{\sigma_{1n}^2}N(u; 0, \sigma_{1n})I(u < 0) + (1 - \pi_{1n}) \cdot \frac{2u^2}{\sigma_{1p}^2}N(u; 0, \sigma_{1p})I(u > 0) \label{eq:wm_alt}\\
  &f_G(u|\phi) = \pi_0 f_{G,0}(u | \sigma_0) + (1-\pi_0)f_{G,1}(u | \pi_{1n},\sigma_{1n}, \sigma_{1p})\label{eq:wm}
\end{align}
This model is fit to the observed test statistics with the Expectation-Maximization (EM) algorithm \citep{dempster1977maximum}, yielding estimates $\hat \phi$. 

Conditional on $\hat\phi$, synthetic datasets $\{\mathbf{u}^n, \mbox{\textbf{fdr}}^n, \mathbf{l}^n\}$ are generated for $n = 1,\dots, N$, where label $l_i$ indicates a not-null test, $l_i = 1-\mathbf{1}\{H_{0i}\}$. First, hypothesis labels $l_i$ are sampled using the estimated proportion $\widehat{\pi}_0$ of null hypotheses (Equation \ref{eq:sim_l}). Test statistics are then generated from their label-specific distributions $f_{G,0}$ or $f_{G,1}$ (Equation \ref{eq:sim_u}). Ground truth local false discovery rates are finally computed using Bayes rule (Equation \ref{eq:comp_fdr}). Each synthetic dataset has the same number of hypotheses as the observed dataset, $i = 1,...,I$.
\begin{align}
  l_i^n &\sim \mbox{Bernoulli}(1-\widehat{\pi}_0) \label{eq:sim_l}\\
  u_i^n &\sim \begin{cases}
    f_{G,0}(u|\hat{\phi})&\text{ if } l_i^n = 0\\
    f_{G,1}(u|\hat{\phi})&\text{ if } l_i^n = 1
  \end{cases}\label{eq:sim_u}\\
  \mbox{fdr}_i^n &= \mbox{fdr}(u_i^n) = \frac{\widehat{\pi}_0f_{G,0}(u_i^n|\hat{\phi})}{f_G(u_i^n|\hat{\phi})} \label{eq:comp_fdr}
\end{align}

\subsection{Step 2: Model Subset Selection}

The synthetic datasets from Step 1 are used to estimate model performances in \textbf{model subset selection}. An fdr model is indexed by $\theta$, a combination of the software package and its user-specified parameter values. Evaluating model $\theta$ given the observed test statistics $\mathbf{u}$ returns an estimate of proportion null and estimates of fdrs, denoted by $\widehat \pi_{0 \theta}(\mathbf{u})$ and $\widehat{\mbox{\textbf{fdr}}}_{\theta}(\mathbf u)$, respectively. Performance is measured by a loss function $L$ that maps model predictions $\widehat{\mbox{\textbf{fdr}}}_\theta(\mathbf u)$ and true $\mbox{\textbf{fdr}}$ to a real number representing prediction error or the information lost. Our loss of interest is mean squared error (MSE): 
\begin{align}
    &L(\widehat{\mbox{\textbf{fdr}}}_\theta(\mathbf u), \mbox{\textbf{fdr}}) = \frac{1}{I}
    ||\widehat{\mbox{\textbf{fdr}}}_\theta(\mathbf u) - \mbox{\textbf{fdr}}||^2 =\frac{1}{I}\sum_{i = 1}^I(\widehat{\mbox{{fdr}}}_{\theta, i}(\mathbf u) - \text{fdr}_i)^2. \label{eq:l}
\end{align}
The goal of model selection is to minimize the objective, $\mathcal{L} = E_{\mathcal{P}}[L(\widehat{\mbox{\textbf{fdr}}}(\mathbf u), \mbox{\textbf{fdr}})]$. The \textbf{fdrSAFE objective estimator}, $\widehat{\mathcal{L}}_{\text{SAFE}}$, is the average of losses from the $N$ synthetic datasets:
  \begin{equation}
  \label{eq:lhat}
  \begin{aligned}
    &\widehat{\mathcal{L}}_{\text{SAFE}}(\theta) = \frac{1}{N}\sum_{n = 1}^NL(\widehat{\mbox{\textbf{fdr}}}_\theta(\mathbf{u}^n), \mbox{\textbf{fdr}}^n).
  \end{aligned}
  \end{equation}

This estimate is unbiased if the synthetic generator is the same as the true data generating model (Section \ref{sec:bias}).
  
We then order models such that $\hat{\mathcal{L}}_{\text{SAFE}}(\theta_{(1)}) \le \dots \le \hat{\mathcal{L}}_{\text{SAFE}}(\theta_{(M)})$ and choose a subset of $m$ models with the best estimated objectives, $\Theta^* = \{\theta_{(1)},\dots,\theta_{(m)}\}$. We use $N=10$ synthetic datasets and $m=10$ selected models based on experiments that can be found in Supplementary Section A.3, though these can be user-specified in our R package.

\subsection{Step 3: Ensemble Aggregation}

The $m$ models with the best estimated performances, $\Theta^*$, are combined in an \textbf{ensemble}. Models with lower (better) estimated objectives are given higher weights. Because local false discovery rates take values in $[0,1]$, their mean squared errors also lie in $[0,1]$, ensuring all weights are nonnegative.
  \begin{align}
    w_{\theta} &= \frac{1 -\widehat{\mathcal{L}}_{\text{SAFE}}(\theta)}{\sum_{\theta' \in \Theta^*} \left(1 -\widehat{\mathcal{L}}_{\text{SAFE}}(\theta')\right)}\\
    \widehat{\pi}_{0,\mbox{{\scriptsize SAFE}}}(\mathbf{u}) &= \sum_{\theta \in \Theta^*} w_{\theta} \widehat{\pi}_{0, \theta}(\mathbf{u})\\
    \widehat{\mbox{\textbf{fdr}}}_{\mbox{{\scriptsize SAFE}}}(\mathbf{u}) &= \sum_{\theta \in \Theta^*} w_{\theta} \widehat{\mbox{\textbf{fdr}}}_\theta(\mathbf{u})
  \end{align}

The ensemble helps to account for uncertainty in the estimated objectives, in contrast to relying only on the single model with minimum estimated objective. We additionally estimate $\widehat{\mbox{\textbf{Fdr}}}_{\text{SAFE}}(\mathbf u)$ using Equation \ref{eq:Fdr_fdr_rel}.

\subsection{Statistical Properties of the fdrSAFE Objective Estimator}
\label{sec:prop}

\subsubsection{Bias}
\label{sec:bias}

In this section, we derive an upper bound for the absolute bias of the fdrSAFE objective estimator. As before, let the subscript $G$ refer to our synthetic generator with joint density of statistics $f_G(\mathbf u)$ while $\mathcal{P}$ refers to the real (unknown) data generating model with joint density of statistics $f_{\mathcal P}(\mathbf u)$. Recall that the function $L(\cdot,\cdot)$ refers to the scalar average loss between two vectors (Equation \ref{eq:l}).

All synthetic datasets are drawn from the same synthetic generator, thus, for a given $\theta$, $L(\widehat{\mbox{\textbf{fdr}}}_\theta(\mathbf{u}^n), \mbox{\textbf{fdr}}^n)$ has the same expectation regardless of $n$. Then,
\begin{align}
  \mathcal{L}(\theta)&= E_{\mathcal{P}}\left[ L(\widehat{\mbox{\textbf{fdr}}}_\theta(\mathbf{u}), \mbox{\textbf{fdr}}) \right] \\
  E\left[\widehat{\mathcal{L}}_{\text{SAFE}}(\theta)\right] &= E_G[L(\widehat{\mbox{\textbf{fdr}}}_\theta(\mathbf{u}), \mbox{\textbf{fdr}})] = E_{\mathcal{P}}\left[\frac{f_G(\mathbf u)}{f_{\mathcal P}(\mathbf u)}L(\widehat{\mbox{\textbf{fdr}}}_\theta(\mathbf{u}), \mbox{\textbf{fdr}})\right] \label{eq:expectation}\\
  \text{Bias} \left[\widehat{\mathcal{L}}_{\text{SAFE}}(\theta)\right] &= E\left[\widehat{\mathcal{L}}_{\text{SAFE}}(\theta)\right] - \mathcal{L}(\theta) = E_{\mathcal{P}}\left[\left(\frac{f_G(\mathbf u)}{f_{\mathcal P}(\mathbf u)}-1\right)L(\widehat{\mbox{\textbf{fdr}}}_\theta(\mathbf{u}), \mbox{\textbf{fdr}})\right] \label{eq:bias}
\end{align}

Equation \ref{eq:expectation} utilizes the fact that the synthetic generator $G$ has the same support as $\mathcal{P}$. We can immediately recognize that if our synthetic generator is equivalent to the real data model, then $\frac{f_G(\mathbf u)}{f_{\mathcal P}(\mathbf u)} = 1$ and the fdrSAFE objective estimator is unbiased. Otherwise, the Cauchy-Schwarz inequality can be used to put an upper bound on absolute bias
  \begin{equation}
  \begin{aligned}
    \big|\mbox{Bias}\big[&\widehat{\mathcal{L}}_{\text{SAFE}}(\theta)\big]\big| \le \Big( \underbrace{E_{\mathcal{P}}\Big[\Big(\frac{f_G(\mathbf u)}{f_{\mathcal P}(\mathbf u)} - 1\Big)^2\Big]}_{(1)} &\underbrace{E_{\mathcal{P}}\Big[L\left(\widehat{\mbox{\textbf{fdr}}}_\theta(\mathbf{u}), \mbox{\textbf{fdr}}\right)^2\Big]}_{(2)}\Big)^{1/2}
  \end{aligned}
  \end{equation}
where (1) $= \mbox{Div}_{\chi^2}(f_G(\mathbf u)||f_{\mathcal P}(\mathbf u))$, the Pearson $\chi^2$-divergence \citep{cressie1984multinomial} between the joint distributions of the dataset under the generator and the true model, and (2) $= \mbox{Var}_{\mathcal{P}}\left[ L\left(\widehat{\mbox{\textbf{fdr}}}_\theta(\mathbf{u}), \mbox{\textbf{fdr}}\right)\right ] + \mathcal{L}(\theta)^2$, the variance of the loss under the real data distribution plus the true objective squared.

\subsubsection{Variance}

In this section, we look at the variance of the fdrSAFE objective estimator. Synthetic datasets are independent from one another and are drawn from the same synthetic generator distribution. The variance depends only on the variance of the loss under the synthetic generator:
\begin{align}
\mbox{Var}&\left[\widehat{\mathcal{L}}_{\text{SAFE}}(\theta)\right] = \frac{1}{N^2}\sum_{n = 1}^N\mbox{Var}_G\left[L(\widehat{\mbox{\textbf{fdr}}}_\theta(\mathbf{u}^n), \mbox{\textbf{fdr}}^n)\right] = \frac{1}{N} \mbox{Var}_G\left[L(\widehat{\mbox{\textbf{fdr}}}_\theta(\mathbf{u}), \mbox{\textbf{fdr}})\right].
\end{align}

\section{Simulation Studies}\label{sec:simulations}

\subsection{Baseline and Ablation Models}
\label{sec:baselines}

As baseline models, we use the three R packages locfdr, fdrtool, and qvalue, each with their default parameter values. Without existing guidance on model selection, these baselines reflect common choices in practice.

To investigate fdrSAFE's performance gain due to ensembling versus model selection, we investigate three ablations. fdrSAFE$_{\text{selection-only}}$ is the single model with the best estimated objective, only utilizing model selection. fdrSAFE$_{\text{aggregation-only}}$ is an equally-weighted ensemble over $m$ randomly chosen models, only utilizing aggregation and with the same ensemble size as fdrSAFE. fdrSAFE$_{\text{aggregation-all}}$ is an equally weighted ensemble over all models in the grid (up to 292).

Finally, we include two oracle approaches. The oracle ensemble is an ensemble of the $m$ models with the best loss using ground truth as known. The single model oracle is the model with the best true loss. These are only applicable in simulations and never an option in application.

\subsection{Simulated Data}
\begin{figure*}
  \centering
  \includegraphics[width = \linewidth]{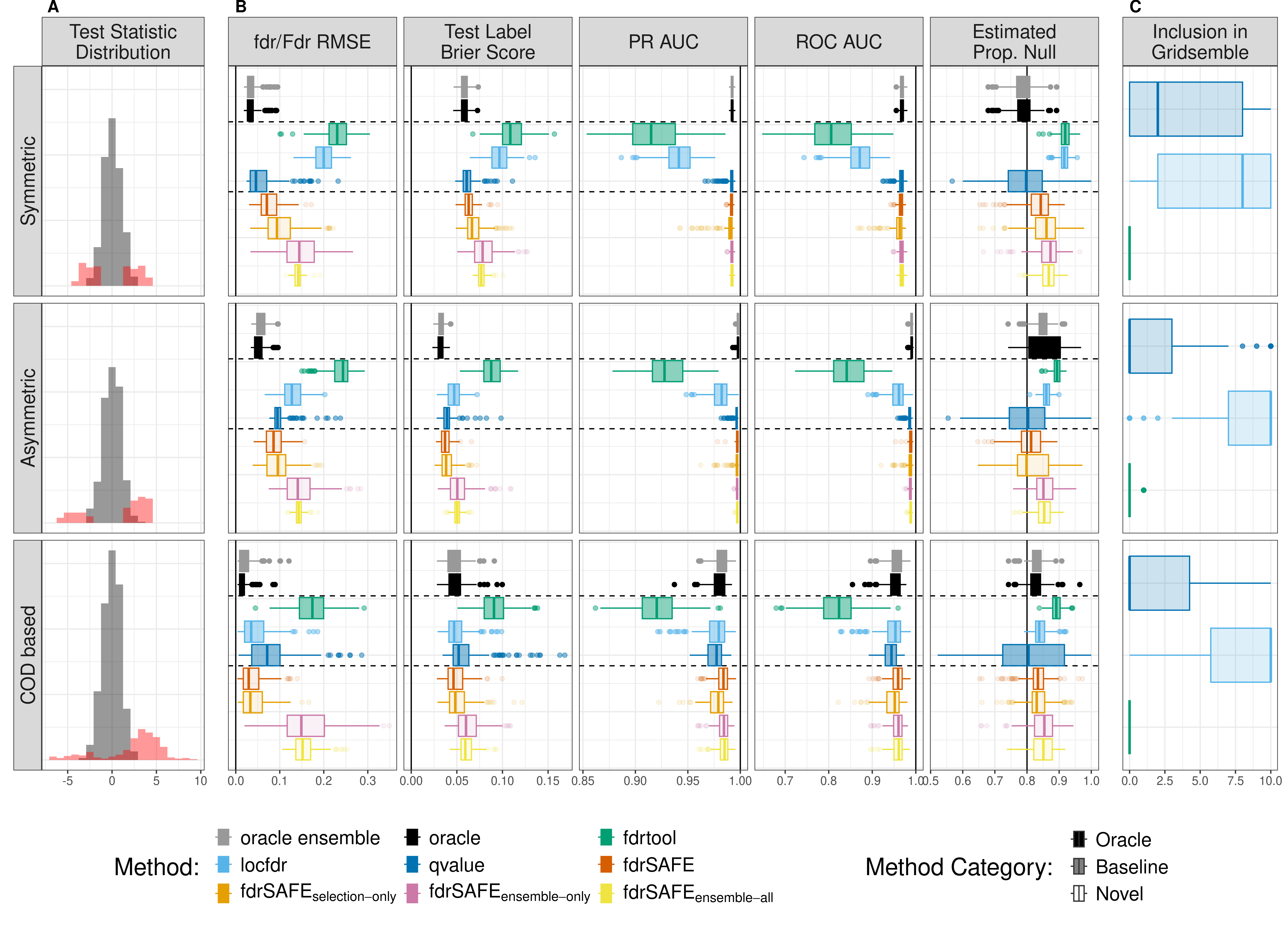}
  
  \caption{\textbf{Results for symmetric, asymmetric and curated ovarian data-based simulation studies.} \textbf{A)} \textbf{Test statistic distributions.} A single simulated example with null tests colored grey, alternative tests colored red. \textbf{B)} \textbf{Model performance metrics} of the oracle and oracle ensemble (above dashed lines), each of the baselines locfdr, fdrtool, and qvalue with their default parameters (between dashed lines), and fdrSAFE including ablations fdrSAFE$_{\text{selection-only}}$, fdrSAFE$_{\text{aggregation-only}}$, and fdrSAFE$_{\text{aggregation-all}}$ (below dashed lines). Plotted are metrics of interest (fdr or Fdr RMSE, test label Brier score, and areas under the PR and ROC curves) and estimated $\pi_0$ over 200 repetitions. Optimal values are indicated with black vertical lines. For exact values of medians, see Supplementary Table B.1. \textbf{C) Proportion of models from each package included in fdrSAFE} over 200 repetitions (not necessarily with default parameters).}
  \label{fig:simulation_results}
\end{figure*}

We consider three simulation studies to evaluate the performance of fdrSAFE, its ablations, and baselines on different types of data. For each, we simulate $200$ repetitions. In all simulations we assume 80\% of hypotheses are null ($\pi_0 = 0.8$). Figure \ref{fig:simulation_results}A visualizes test statistics generated from each study.

The first two simulation settings intentionally differ from fdrSAFE’s synthetic generator. This allows us to assess the performance of fdrSAFE when the synthetic generator does not reflect the true data-generating mechanism.

\underline{Symmetric}: Based on \cite{strimmer}, statistics from null hypotheses follow a Normal distribution with mean zero and those from alternative hypotheses follow a mixture of Uniforms away from zero. Each repetition consists of $I=1000$ hypotheses. Specifically:
\begin{align*}
  f_0 &= N(\mu = 0, \sigma = 1)\\
  f_1 &= \frac{1}{2}U(-4, -1.33) + \frac{1}{2}U(1.33, 4).
\end{align*}
The true null distribution is used to convert test-statistics to p-values for qvalue.
 
\underline{Asymmetric}: In the asymmetric study, the majority of alternative hypotheses have positive test statistics, but the range of negative statistics is wider. Each repetition consists of $I=1000$ hypotheses. Specifically:
\begin{align*}
  f_0 &= N(\mu = 0, \sigma = 1)\\
  f_1 &= \frac{1}{3}U(-6, -2.5) + \frac{2}{3}U(1.5, 4.5).
\end{align*}
Again, the true null distribution is used to convert test-statistics to p-values for qvalue. We note that in this setting, the conversion to p-values loses information regarding asymmetry, which may be a drawback of qvalue.

\underline{Curated Ovarian Data (COD) Based}: This simulation study is based on real gene expression data from the {\tt curatedOvarianData} R software package \citep{curatedovarian}. We start with the TCGA microarray expression set of 13,104 probes across 578 samples and subset to the top 10\% ($I = 1311$) most variable probes. Empirical mean $\hat \mu$ and covariance $\hat \Sigma$ are computed directly from the data to maintain a realistic scale and correlation structure. 

We simulate expression for 20 samples, with 10 from each of two groups, A and B. Each sample's expression is simulated following a multivariate normal with its respective group mean and shared covariance $\hat \Sigma$ (Equation \ref{eq:cod_exp}). The mean for group A, $\mu_A$, is set to the empirical $\hat \mu$. The mean for group B, $\mu_B$ is set at $\mu_A$ for probes with no differential expression (null tests), and offset by a term $\delta$ for probes with differential expression (Equations \ref{eq:cod_means}, \ref{eq:cod_mag}). Of the probes with differential expression, 20\% will show underexpression in group B while 80\% show overexpression in group B (Equation \ref{eq:cod_asymmetry}). Finally, probe-wise two-sample t-tests are used to compute test statistics (Equation \ref{eq:cod_test}). In summary:
\begin{align}
  \text{Expression }&\quad X_A \sim MVN(\mu_A, \hat\Sigma) \in \mathbb R^{I \times 10}, \quad\quad X_B \sim MVN(\mu_B, \hat\Sigma) \in \mathbb R^{I \times 10} \label{eq:cod_exp}\\
  &\quad \mu_{A} = \hat \mu, \quad\quad \mu_{B,i} = \begin{cases}
    \mu_{A, i} + d_i\cdot \delta_i & \text{ if } l_i = 1\\
    \mu_{A,i} & \text{ if } l_i = 0
  \end{cases} \label{eq:cod_means}\\
  \text{Offset Magnitude}&\quad\delta_i \sim N(2, \sigma = 0.5) \label{eq:cod_mag}\\
  \text{Offset Direction}&\quad p(d_i) = 0.2\cdot \mathbf 1(d_i = -1) + 0.8\cdot \mathbf 1(d_i = 1) \label{eq:cod_asymmetry}\\
  \text{Test Statistic}&\quad u_i = \left(\bar X_{i,B} - \bar X_{i,A} \right) / \sqrt{s^2_{\bar X_{i,B}} + s^2_{\bar X_{i,A}}} \label{eq:cod_test}
\end{align}
The null distribution $t_{(18)}$ is used to convert test-statistics to p-values for the qvalue package. While such t-distributions are commonly used in practice, it is not technically appropriate in this case due to correlated gene expression, and thus, correlated test-statistics. This violation of model assumptions may prove problematic for qvalue.

\subsection{Evaluation}
\label{sec:eval}
The first metric to evaluate fdr estimation approaches is the \underline{estimated proportion null} $\hat \pi_{0\theta}(\mathbf u)$, which is compared to the true value $\pi_0 = 0.8$. 

The next two metrics are important for the goal of accurate Fdr and fdr estimates. \underline{fdr RMSE} is the root mean squared error between the estimated and true local fdr values. The true fdr can be calculated only for symmetric and asymmetric studies using the known distributions of test statistics. For the COD-based study, we instead look at \underline{Fdr RMSE} as the RMSE between estimated and empirical tail-end Fdr values.

The last three metrics are important for the goal of accurate classifications of hypotheses into null versus alternative. \underline{PR AUC} and \underline{ROC AUC} are the areas under the precision-recall and receiver operating characteristic curves, respectively. \underline{Test label Brier score} is the average squared distance between the positive predictive values, $1-\widehat{\mbox{\textbf{fdr}}}_{\theta}(\mathbf u)$, and binary test labels, $\mathbf{l}$.

\subsection{Results}
\label{sec:results}

Figure \ref{fig:simulation_results}B reports the five metrics for fdrSAFE along with all baseline, ablation, and oracle methods. The key result from these simulations is that fdrSAFE demonstrates robust near-optimality, performing comparably to or better than the best baseline in all settings. Its performance does not deteriorate as asymmetry is introduced or independence assumptions are violated. This is not to say that it is \textit{always} the best, but it adapts to the observed data to perform well across a range of settings.

The best baseline method seems to be qvalue, but its performance declines as asymmetry is introduced and assumptions are not met. While locfdr is particularly strong in the latter, more complex settings, it performs poorly in the simplest, symmetric simulation study. In all settings we evaluated, fdrtool shows poor performance --- more analysis is required to understand where fdrtool performs best. 

In terms of estimated proportion null $\hat\pi_0$, qvalue demonstrates the least bias but also the greatest variability. We see even the oracle approaches overestimate $\pi_0$ in the asymmetric and COD-based settings, which shows it may be unrealistic to expect a correct $\hat\pi_0$ when optimizing for fdr MSE in model subset selection.

fdrSAFE also outperforms each ablation. The improvement of fdrSAFE over fdrSAFE$_{\text{selection-only}}$ and fdrSAFE$_{\text{aggregation-only}}$ emphasizes the utility of combining the model subset selection and ensemble procedures. The improvement of fdrSAFE over fdrSAFE$_{\text{aggregation-all}}$ highlights that being selective over which models to include in an ensemble is more important than increasing ensemble size.

Finally, despite the misspecification of the synthetic generator in both the symmetric and asymmetric settings, fdrSAFE demonstrates strong and consistent performance, suggesting that it remains effective even when the synthetic generator deviates from the true generative process.

\section{Experimental Application}
\label{sec:experiment}

\subsection{Data}
The Platinum Spike dataset \citep{platinum} is an experimental spike-in study of 18 Affymetrix Drosophila Genome 2.0 microarrays under two conditions. A spike-in experiment artificially changes a known set of genes at controlled concentrations, allowing the true differential expression status of each gene to be known in advance. There are three observations for each condition with three technical replicates per observation. Expression data are available on GEO Browser as \href{https://www.ncbi.nlm.nih.gov/geo/query/acc.cgi}{GSE21344} and the fold change data is in an additional file of the Platinum Spike paper, \citet{platinum}. 

Because the spike-in ground truth is defined at the gene level, we summarize the 535,824 Affymetrix probes to 18,952 genes using the robust multi-array average (RMA) method \citep{irizarry2003exploration}. We remove empty probe sets, those assigned to multiple clones, and those whose clones have multiple fold change values. This results in expression measures for $I =$ 5,370 genes, where 36.2\% are differentially expressed ($\pi_0 = 0.638$) with fold changes between 0.25 and 3.5. We average across technical replicates before computing a two-sample t-statistic for each gene.

\subsection{Analysis}

We apply fdrSAFE and all of the baselines to the differential expression test statistics, where p-values for the qvalue package are derived using a $t_{(4)}$ null distribution. To quantify uncertainty around evaluation metrics, we performed bootstrap resampling of genes 1000 times.

In addition to the fdr evaluation metrics, two types of calibration are considered. First, we inspect the calibration of local false discovery rates by comparing, within bins, average predicted probability null versus true proportion null. This measures whether a model’s predicted fdr values are reliable as probabilities. Second, we compare estimated to true global false discovery rate for each unique decision cutoff. This measures whether a method accurately controls the expected proportion of false discoveries among all discoveries made at or below a given fdr threshold.

\begin{figure}
  \centering
  \includegraphics[width=\linewidth]{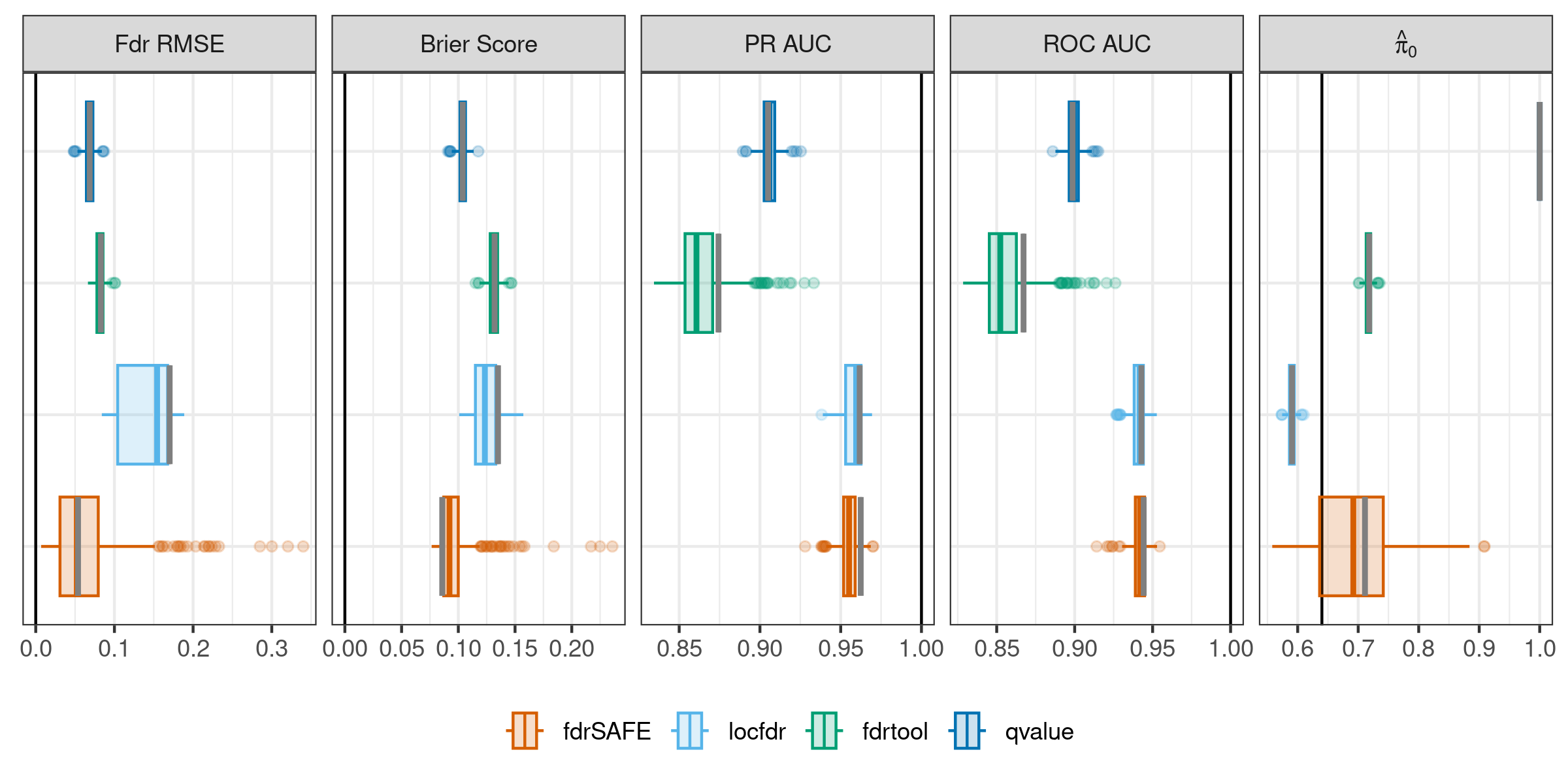}
  \caption{\textbf{Model performances on experimental application}: comparing fdrSAFE to each of the baseline methods across 1000 bootstrap resamplings of genes. Metrics on full dataset reported with gray vertical lines. Ideal values of each metric are indicated with black vertical lines.}
  \label{fig:platinum_boostrap_metrics}
\end{figure}

\begin{figure}
  \centering
  \includegraphics[width = \textwidth]{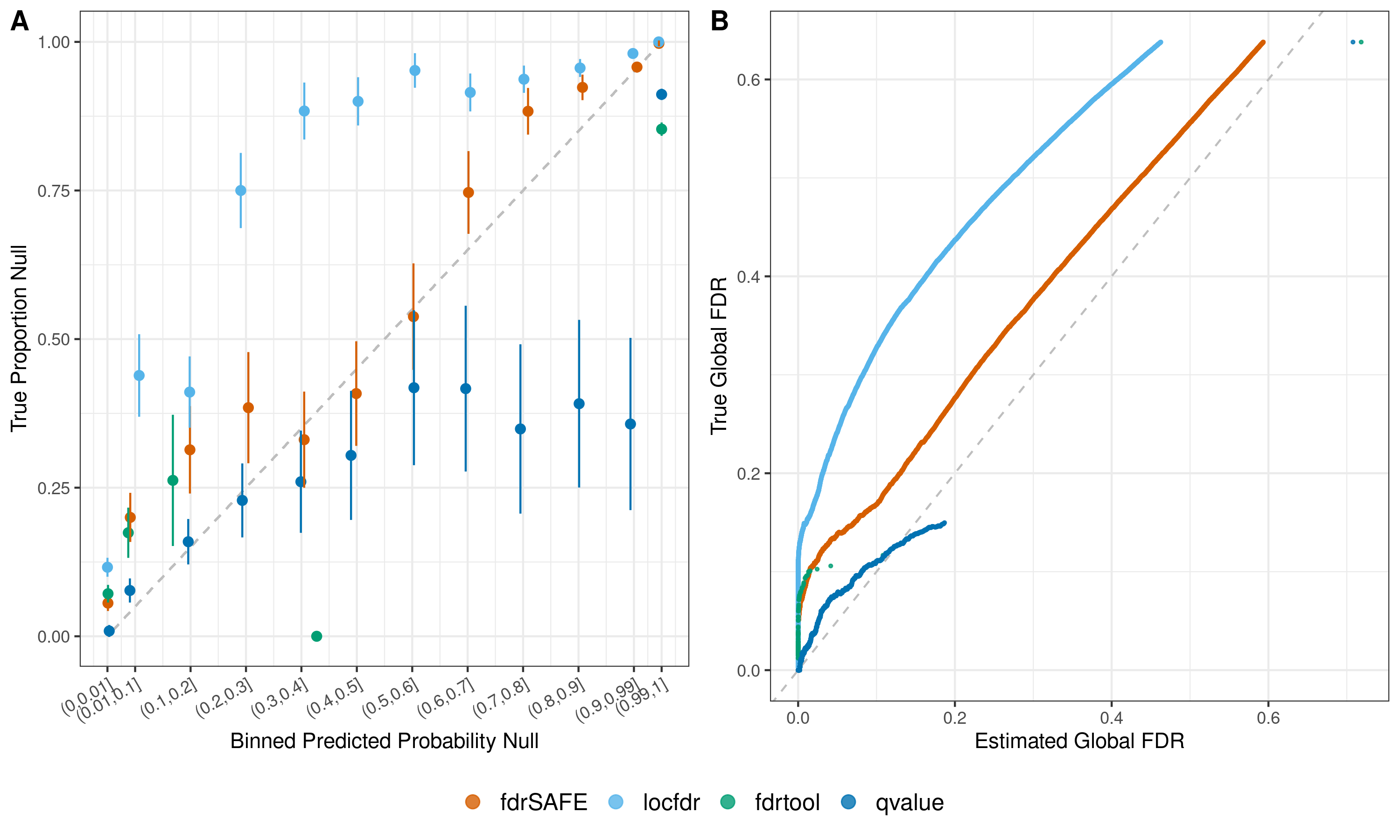}
  \caption{\textbf{Calibration of experimental application results} comparing fdrSAFE to each of the baseline methods. \textbf{A) Calibration of local false discovery rates (fdr).} For each method and bin of fdr values, plotted is the average of fdr values (x-axis) versus true proportion of tests which are null with 95\% binomial confidence intervals (y-axis). Perfect calibration is indicated by the dashed line. Bins are evenly spaced, with the exception of the first bin [0, 0.01] and final bin (0.99, 1], which is to emphasize how many tests have estimated extreme fdr values. \textbf{B) Calibration of global false discovery rate (FDR).} For each method, the set of unique estimated local fdr values is taken as candidate thresholds. At a given threshold, we compute (i) the estimated global FDR: the mean of the estimated local fdr values for all tests below the threshold (x-axis), and (ii) the true global FDR: the fraction of those tests that are truly null (y-axis). Perfect calibration is indicated by the dashed line.}
  \label{fig:platinum_calibration}
\end{figure}

\subsection{Results}
fdrSAFE shows the best results on Fdr RMSE and Brier Score, and is tied with locfdr for PR and ROC AUC (Figure \ref{fig:platinum_boostrap_metrics}). fdrSAFE also outperforms the baselines in terms of estimated proportion null. These results most closely match the COD-based simulation study with locfdr trailing closely behind fdrSAFE.

fdrSAFE also offers a balance of calibration and interpretability. It is the closest to the diagonal in the binned fdr calibration plot (Figure \ref{fig:platinum_calibration}A), indicating good probabilistic calibration. In the global FDR calibration plot (Figure \ref{fig:platinum_calibration}B), fdrSAFE maintains a continuous and well-spread trajectory across the full range of FDR values, unlike qvalue and fdrtool, which collapse most predictions into extreme values. However, qvalue achieves the closest alignment with true global FDR at realistic thresholds ($\le$ 0.05), suggesting it may be preferable in terms of global FDR.

The sharp jump in true global FDR observed in Figure \ref{fig:platinum_calibration}B for qvalue and fdrtool corresponds to a large number of hypotheses with identical or near-identical fdr estimates. In particular, qvalue and fdrtool assign fdr at or above 0.99 to 64\% and 71\% of hypotheses, respectively. This concentration of extreme values causes many discoveries to enter at once when the threshold exceeds 0.99, resulting in a visible discontinuity. By contrast, fdrSAFE and locfdr assign such extreme values to fewer than 10\% of hypotheses, allowing for a smoother global FDR curve.

\section{Discussion}

We designed and evaluated fdrSAFE, a practical and data-driven approach to estimate local false discovery rates. These quantities are valuable for quantifying uncertainty, but their applicability is challenged by the lack of practical guidance for model selection, often leading to arbitrary model choice. fdrSAFE overcomes this problem using a three-step approach: generating synthetic datasets, selecting top models based on performance on synthetic data, and ensembling their estimates. This work is implemented in an open-source R software package on GitHub at \href{https://github.com/jennalandy/fdrSAFE}{jennalandy/fdrSAFE}. 

Importantly, we observe that across simulated and experimental data, fdrSAFE demonstrates robust near-optimality, a quality no baseline method has.
Along with its encouraging performance, we emphasize that without fdrSAFE, practitioners must currently make arbitrary choices of what inference to report, compromising the replicability of results reported in the presence of multiplicities. Beyond the introduction and evaluation of fdrSAFE, this work also provides a systematic evaluation of the three fdr packages under their default configurations, offering insight into their relative strengths and limitations in real-world use. 

While fdrSAFE increases computational cost by fitting multiple fdr models to both synthetic and observed datasets, this cost is modest. All models in the grid are individually fast, and for the number of tests considered in this paper, fdrSAFE completes in well under a minute without parallelization, and faster when multiple cores are available (controlled by the {\tt parallel} and {\tt n\_workers} parameters). For larger datasets, our open-source R package allows users to reduce runtime by setting the size of synthetic datasets to be smaller than the observed data with the {\tt synthetic\_size} parameter.

We see four main avenues for future research with fdrSAFE. First, other distributions or fitting methods could be used for the synthetic generator. For instance, our Normal null distribution could be replaced with a scaled t-distribution to allow for heavier tails. The synthetic generator could also borrow estimates from an existing fdr model. We found, though, that a synthetic generator fit from scratch does as good or better than borrowing estimates from baseline models, especially when baselines severely overestimate $\pi_0$. Fitting from scratch also ensures synthetic datasets are not biased towards the model the estimates were borrowed from. 

The fdrSAFE approach could also be extended to other types of test statistics with adjustments to the synthetic generator and the grid of possible fdr models. The null and alternative components of the synthetic generator should match the assumed low- and high-evidence regions of the test statistic. Each model in the grid must be able to use the test statistic as input, possibly after a conversion, as seen with qvalue here.

We chose to use a grid search for model subset selection, but could have alternatively performed a random search or Bayesian optimization~\citep{yu2020hyper} using our estimator for the objective. We saw no difference in performance when a random search approach was tested for fdrSAFE and all ablations. Bayesian optimization is challenging with categorical parameters \citep{garrido2020dealing}, of which the R packages have many, so we chose not to explore this route. With regard to optimization, the approach underlying fdrSAFE can be implemented with other loss functions if different behavior is desired. For example, one may transform fdr prior to evaluating the mean squared error, such as to emphasize deviations near the extremes, or add additional penalties or regularizations.

Finally, although this work borrows some strategies from model selection, at this time we do not perform train-test-split or cross validation. This is because we would run into the same issue that prompted the development of fdrSAFE itself: fdrs are unobservable, so model performances cannot be estimated empirically on a test set.

Overall, fdrSAFE offers a practical, data-driven solution to the challenge of selecting among competing local false discovery rate estimation methods without access to ground truth. By leveraging synthetic data and selective ensembling, it reliably estimates local fdr while improving reproducibility in large-scale multiple testing. 

\section*{Acknowledgements}
Jenna Landy was supported by the NIH-NIGMS training grant T32GM135117 and NIH-NCI grant 5R01 CA262710-03. Giovanni Parmigiani was supported by NSF-DMS grants 1810829 and 2113707 and NIH-NCI grant 5R01 CA262710-03.

\bibliographystyle{unsrtnat}
\bibliography{references}

\begin{thebibliography}{34}
\providecommand{\natexlab}[1]{#1}
\providecommand{\url}[1]{\texttt{#1}}
\expandafter\ifx\csname urlstyle\endcsname\relax
  \providecommand{\doi}[1]{doi: #1}\else
  \providecommand{\doi}{doi: \begingroup \urlstyle{rm}\Url}\fi

\bibitem[Dudoit et~al.(2003)Dudoit, Shaffer, and Boldrick]{Dudoit2003}
Sandrine Dudoit, Juliet~Popper Shaffer, and Jennifer~C. Boldrick.
\newblock Multiple hypothesis testing in microarray experiments.
\newblock \emph{Statistical Science}, 18\penalty0 (1):\penalty0 71--103, 2003.
\newblock \doi{10.1214/ss/1056397487}.

\bibitem[Farcomeni(2008)]{farcomeni2008review}
Alessio Farcomeni.
\newblock A review of modern multiple hypothesis testing, with particular attention to the false discovery proportion.
\newblock \emph{Statistical Methods in Medical Research}, 17\penalty0 (4):\penalty0 347--388, 2008.
\newblock \doi{10.1177/0962280206079046}.

\bibitem[Efron et~al.(2001)Efron, Tibshirani, Storey, and Tusher]{efron2001}
Bradley Efron, Robert Tibshirani, John~D Storey, and Virginia Tusher.
\newblock Empirical bayes analysis of a microarray experiment.
\newblock \emph{Journal of the American Statistical Association}, 96\penalty0 (456):\penalty0 1151--1160, 2001.
\newblock \doi{10.1198/016214501753382129}.

\bibitem[Efron et~al.(2015)Efron, Turnbull, Narasimhan, and Strimmer]{locfdr}
Bradley Efron, Brit Turnbull, Balasubramanian Narasimhan, and Korbinian Strimmer.
\newblock \emph{\texttt{locfdr}: Computes Local False Discovery Rates}, 2015.
\newblock R package version 1.1-8.

\bibitem[Klaus and Strimmer(2021)]{fdrtool}
Bernd Klaus and Korbinian Strimmer.
\newblock \emph{\texttt{fdrtool}: Estimation of (Local) False Discovery Rates and Higher Criticism}, 2021.
\newblock R package version 1.2.17.

\bibitem[Strimmer(2008)]{strimmer}
Korbinian Strimmer.
\newblock fdrtool: a versatile r package for estimating local and tail area-based false discovery rates.
\newblock \emph{Bioinformatics}, 24\penalty0 (12):\penalty0 1461--1462, 2008.
\newblock \doi{10.1093/bioinformatics/btn209}.

\bibitem[Storey et~al.(2015)Storey, Bass, Dabney, and Robinson]{qvalue}
John Storey, Andrew Bass, Alan Dabney, and David Robinson.
\newblock \emph{\texttt{qvalue}: Q-value estimation for false discovery rate control}, 2015.
\newblock R package version 2.28.0.

\bibitem[Tukey(1953)]{tukey1953problem}
John~W. Tukey.
\newblock The problem of multiple comparisons.
\newblock Technical report, Princeton University, 1953.

\bibitem[Tukey(1991)]{Tukey:1991ga}
John~W. Tukey.
\newblock The philosophy of multiple comparisons.
\newblock \emph{Statistical Science}, pages 100--116, 1991.
\newblock \doi{10.1214/ss/1177011945}.

\bibitem[Benjamini and Hochberg(1995)]{benj.hoch}
Y.~Benjamini and Y.~Hochberg.
\newblock Controlling the false discovery rate: A practical and powerful approach to multiple testing.
\newblock \emph{Journal of the Royal Statistical Society Series B (Methodological)}, 57:\penalty0 289--300, 1995.
\newblock \doi{10.1111/j.2517-6161.1995.tb02031.x}.

\bibitem[Efron and Tibshirani(2002)]{efro:tibs:2002}
Bradley Efron and Robert Tibshirani.
\newblock {Empirical {B}ayes methods and false discovery rates for microarrays.}
\newblock \emph{Genetic Epidemiology}, 23\penalty0 (1):\penalty0 70--86, June 2002.
\newblock \doi{10.1002/gepi.1124}.

\bibitem[Efron(2007)]{efron2007}
Bradley Efron.
\newblock Size, power and false discovery rates.
\newblock \emph{The Annals of Statistics}, 35:\penalty0 1351--1377, 2007.
\newblock \doi{10.1214/009053606000001460}.

\bibitem[Stephens(2016)]{Stephens:2017en}
Matthew Stephens.
\newblock {False discovery rates: a new deal}.
\newblock \emph{Biostatistics}, 18\penalty0 (2):\penalty0 275--294, April 2016.
\newblock ISSN 1465-4644.
\newblock \doi{10.1093/biostatistics/kxw041}.

\bibitem[M\"uller et~al.(2007)M\"uller, Parmigiani, and Rice]{muel:parm:rice:2007}
Peter M\"uller, Giovanni Parmigiani, and Kenneth Rice.
\newblock In J.~M. Bernardo, M.~J. Bayarri, J.~O. Berger, A.~P. Dawid, D.~Heckerman, A.~F.~M. Smith, and M.~West, editors, \emph{Bayesian Statistics 8}. Oxford University Press, 2007.
\newblock \doi{10.1093/oso/9780199214655.003.0014}.

\bibitem[Yu and Zhu(2020)]{yu2020hyper}
Tong Yu and Hong Zhu.
\newblock Hyper-parameter optimization: A review of algorithms and applications.
\newblock \emph{arXiv preprint arXiv:2003.05689}, 2020.
\newblock \doi{10.48550/arXiv.2003.05689}.

\bibitem[Kuhn et~al.(2013)Kuhn, Johnson, et~al.]{kuhn2013applied}
Max Kuhn, Kjell Johnson, et~al.
\newblock \emph{Applied predictive modeling}, volume~26.
\newblock Springer, 2013.
\newblock \doi{10.1007/978-1-4614-6849-3}.

\bibitem[Nomura and Saito(2021)]{nomura2021}
Masahiro Nomura and Yuta Saito.
\newblock Efficient hyperparameter optimization under multi-source covariate shift.
\newblock \emph{CIKM}, 2021.
\newblock \doi{10.1145/3459637.3482336}.

\bibitem[Efron(1982)]{efron1982jackknife}
Bradley Efron.
\newblock \emph{The jackknife, the bootstrap and other resampling plans}.
\newblock SIAM, 1982.

\bibitem[Kohavi et~al.(1995)]{kohavi1995study}
Ron Kohavi et~al.
\newblock A study of cross-validation and bootstrap for accuracy estimation and model selection.
\newblock In \emph{Ijcai}, volume~14, pages 1137--1145. Montreal, Canada, 1995.

\bibitem[Dong et~al.(2020)Dong, Yu, Cao, Shi, and Ma]{Dong2020}
Xibin Dong, Zhiwen Yu, Wenming Cao, Yifan Shi, and Qianli Ma.
\newblock A survey on ensemble learning.
\newblock \emph{Frontiers of Computer Science}, 14\penalty0 (2):\penalty0 241--258, 2020.
\newblock ISSN 2095-2236.
\newblock \doi{10.1007/s11704-019-8208-z}.

\bibitem[Madigan and Raftery(1994)]{MadiganRaftery1994}
David Madigan and Adrian~E. Raftery.
\newblock Model selection and accounting for model uncertainty in graphical models using {O}ccam's {W}indow.
\newblock \emph{Journal of the American Statistical Association}, 89\penalty0 (428):\penalty0 1535--1546, 1994.
\newblock ISSN 01621459.
\newblock \doi{10.1080/01621459.1994.10476894}.

\bibitem[Ryu et~al.(2002)Ryu, Jones, Blades, Parmigiani, Hollingsworth, Hruban, and Kern]{ryu2002relationships}
Byungwoo Ryu, Jessa Jones, Natalie~J Blades, Giovanni Parmigiani, Michael~A Hollingsworth, Ralph~H Hruban, and Scott~E Kern.
\newblock Relationships and differentially expressed genes among pancreatic cancers examined by large-scale serial analysis of gene expression.
\newblock \emph{Cancer Research}, 62\penalty0 (3):\penalty0 819--826, 2002.

\bibitem[Cui and Churchill(2003)]{cui2003statistical}
Xiangqin Cui and Gary~A Churchill.
\newblock Statistical tests for differential expression in c{DNA} microarray experiments.
\newblock \emph{Genome Biology}, 4:\penalty0 1--10, 2003.
\newblock \doi{10.1186/gb-2003-4-4-210}.

\bibitem[Lu et~al.(2005)Lu, Tomfohr, and Kepler]{lu2005identifying}
Jun Lu, John~K Tomfohr, and Thomas~B Kepler.
\newblock Identifying differential expression in multiple sage libraries: an overdispersed log-linear model approach.
\newblock \emph{BMC Bioinformatics}, 6:\penalty0 1--14, 2005.
\newblock \doi{10.1186/1471-2105-6-165}.

\bibitem[Robinson and Smyth(2007)]{robinson2007moderated}
Mark~D Robinson and Gordon~K Smyth.
\newblock Moderated statistical tests for assessing differences in tag abundance.
\newblock \emph{Bioinformatics}, 23\penalty0 (21):\penalty0 2881--2887, 2007.
\newblock \doi{10.1093/bioinformatics/btm453}.

\bibitem[Chen et~al.(2025)Chen, Chen, Lun, Baldoni, and Smyth]{chen2020edger}
Y~Chen, L~Chen, ATL Lun, P~Baldoni, and GK~Smyth.
\newblock edge{R} v4: powerful differential analysis of sequencing data with expanded functionality and improved support for small counts and larger datasets.
\newblock \emph{Nucleic Acids Research}, 53\penalty0 (2), 2025.
\newblock \doi{10.1093/nar/gkaf018}.

\bibitem[Varet et~al.(2016)Varet, Brillet-Gu{\'e}guen, Copp{\'e}e, and Dillies]{varet2016sartools}
Hugo Varet, Loraine Brillet-Gu{\'e}guen, Jean-Yves Copp{\'e}e, and Marie-Agn{\`e}s Dillies.
\newblock {SART}ools: a {DES}eq2-and edge{R}-based {R} pipeline for comprehensive differential analysis of {RNA-S}eq data.
\newblock \emph{PLOS One}, 11\penalty0 (6):\penalty0 e0157022, 2016.
\newblock \doi{10.1371/journal.pone.0157022}.

\bibitem[Rossell and Telesca(2017)]{Rossell2017}
David Rossell and Donatello Telesca.
\newblock Non-local priors for high-dimensional estimation.
\newblock \emph{Journal of the American Statistical Association}, 112:\penalty0 254--265, 2017.
\newblock \doi{10.1080/01621459.2015.1130634}.

\bibitem[Dempster et~al.(1977)Dempster, Laird, and Rubin]{dempster1977maximum}
Arthur~P Dempster, Nan~M Laird, and Donald~B Rubin.
\newblock Maximum likelihood from incomplete data via the em algorithm.
\newblock \emph{Journal of the Royal Statistical Society: Series B (Methodological)}, 39\penalty0 (1):\penalty0 1--22, 1977.
\newblock \doi{10.1111/j.2517-6161.1977.tb01600.x}.

\bibitem[Cressie and Read(1984)]{cressie1984multinomial}
Noel Cressie and Timothy~RC Read.
\newblock Multinomial goodness-of-fit tests.
\newblock \emph{Journal of the Royal Statistical Society Series B: Statistical Methodology}, 46\penalty0 (3):\penalty0 440--464, 1984.

\bibitem[Ganzfried et~al.(2013)Ganzfried, Riester, Haibe-Kains, Risch, Tyekucheva, Jazic, Wang, Ahmadifar, Birrer, Parmigiani, Huttenhower, and Waldron]{curatedovarian}
BF~Ganzfried, M~Riester, B~Haibe-Kains, T~Risch, S~Tyekucheva, I~Jazic, XV~Wang, M~Ahmadifar, M~Birrer, G~Parmigiani, C~Huttenhower, and L~Waldron.
\newblock curated{O}varian{D}ata: Clinically annotated data for the ovarian cancer transcriptome.
\newblock \emph{Database}, 2013.
\newblock \doi{10.1093/database/bat013}.

\bibitem[Zhu et~al.(2010)Zhu, Miecznikowski, and Halfon]{platinum}
Q.~Zhu, J.C. Miecznikowski, and M.S. Halfon.
\newblock Preferred analysis methods for affymetrix genechips. {II}. an expanded, balanced, wholly-defined spike-in dataset.
\newblock \emph{BMC Bioinformatics}, 11:\penalty0 285, 2010.
\newblock \doi{10.1186/1471-2105-11-285}.

\bibitem[Irizarry et~al.(2003)Irizarry, Hobbs, Collin, Beazer-Barclay, Antonellis, Scherf, and Speed]{irizarry2003exploration}
Rafael~A Irizarry, Bridget Hobbs, Francois Collin, Yasmin~D Beazer-Barclay, Kristen~J Antonellis, Uwe Scherf, and Terence~P Speed.
\newblock Exploration, normalization, and summaries of high density oligonucleotide array probe level data.
\newblock \emph{Biostatistics}, 4\penalty0 (2):\penalty0 249--264, 2003.
\newblock \doi{10.1093/biostatistics/4.2.249}.

\bibitem[Garrido-Merch{\'a}n and Hern{\'a}ndez-Lobato(2020)]{garrido2020dealing}
Eduardo~C Garrido-Merch{\'a}n and Daniel Hern{\'a}ndez-Lobato.
\newblock Dealing with categorical and integer-valued variables in bayesian optimization with gaussian processes.
\newblock \emph{Neurocomputing}, 380:\penalty0 20--35, 2020.
\newblock \doi{10.1016/j.neucom.2019.11.004}.

\end{thebibliography}

\newpage

\captionsetup[table]{name=Supplementary Table}
\captionsetup[figure]{name=Supplementary Figure}

\renewcommand{\shorttitle}{Supplementary Materials for \textit{fdrSAFE: Selective Aggregation for False Discovery Rates}}

\thispagestyle{plain}

\setcounter{table}{0}
\setcounter{figure}{0}
\setcounter{section}{0}

\renewcommand{\thesection}{\Alph{section}}
\renewcommand{\thefigure}{\Alph{section}.\arabic{figure}}
\renewcommand{\thetable}{\Alph{section}.\arabic{table}}

\begin{center}
    \rule{\textwidth}{2pt}\\
        {\LARGE\sc Supplementary Materials for \\ \textit{fdrSAFE: Selective Aggregation for False Discovery Rates} \par}
    \vspace{0.1in}
    \rule{\textwidth}{2pt}\\
    \vspace{0.2cm}
    \textsc{\undertitle}\\
\end{center}

\section{Methods: Additional Details}

\subsection{Grid of fdr Models}
\label{sec:software}

Local fdr models are indexed by $\theta$, which includes the package and parameter values. A grid of 292 models is used, including all possible combinations of categorical parameters and evenly spaced values of numeric parameters. Because fdrtool only has two parameters while locfdr and qvalue each have four, there are fewer models using the fdrtool package. For a given dataset, models that result in convergence errors on observed data are excluded.

The locfdr package \citep{efron2007} assumes a Normal distribution for the weighted null density $\pi_0 f_0$. We consider four key parameters. \texttt{nulltype} determines whether the weighted null is estimated by maximum likelihood or central matching. \texttt{pct0} determines the quantile range \texttt{[pct0, 1-pct0]} of central test statistics used to fit the weighted null. The default value is 0.25, and in our grid, we consider \texttt{pct0} $\in$ \texttt{[0, 0.3]} (using 40\%-100\% of the data to fit the null). \texttt{type} determines whether the marginal density $f$ is fit with a natural spline or polynomial model. \texttt{pct} is the proportion of outliers excluded when fitting the marginal density. The default value is 0, and in our grid, we consider \texttt{pct} $\in$ \texttt{[0, 0.3]} (using 70\%-100\% of the data to fit the marginal). Out of the full grid of 292 models, 150 use locfdr.

The fdrtool package \citep{strimmer, fdrtool} also assumes a Normal distribution for the weighted null, but uses a non-parametric Grenander estimator for the marginal density. We consider two key parameters. \texttt{cutoff.method} chooses the method that determines which central test statistics to use when fitting the weighted null. If the \texttt{cutoff.method = "pct0"}, then \texttt{pct0} defines the proportion of central statistics used to fit the null. We consider \texttt{pct0} $\in$ {\tt [0.4, 1]}, matching the range used in locfdr. Out of the full grid of 292 models, 22 use fdrtool.

The qvalue package \citep{qvalue} takes in p-values and estimates the weighted null p-value density as a Uniform distribution. It uses a natural spline to fit the marginal density. We consider four key parameters. \texttt{pi0.method} determines whether to use a bootstrap or a smoother to tune a parameter used to estimate $\pi_0$ (details in Supplementary Section \ref{sec:qvalue_distribution}). \texttt{transf} is whether a probit or logit transformation is applied to the p-values. \texttt{adj} defines the smoothing bandwidth used in the full density estimation. The default value is 1.5, and we consider values of \texttt{adj} $\in$ {\tt [0.5, 2]}. If \texttt{pi0.method = "smoother"}, then \texttt{smooth.log.pi0} controls whether a smoother is applied to $\hbox{log}(\pi_0)$ in the procedure estimating $\pi_0$.  Out of the full grid of 292 models, 120 use qvalue.

\newpage
\subsection{Choosing $\mathbf{N}$ and $\mathbf{m}$}
\label{sec:complex}

\begin{figure}[H]
    \centering
    \includegraphics[width = \linewidth]{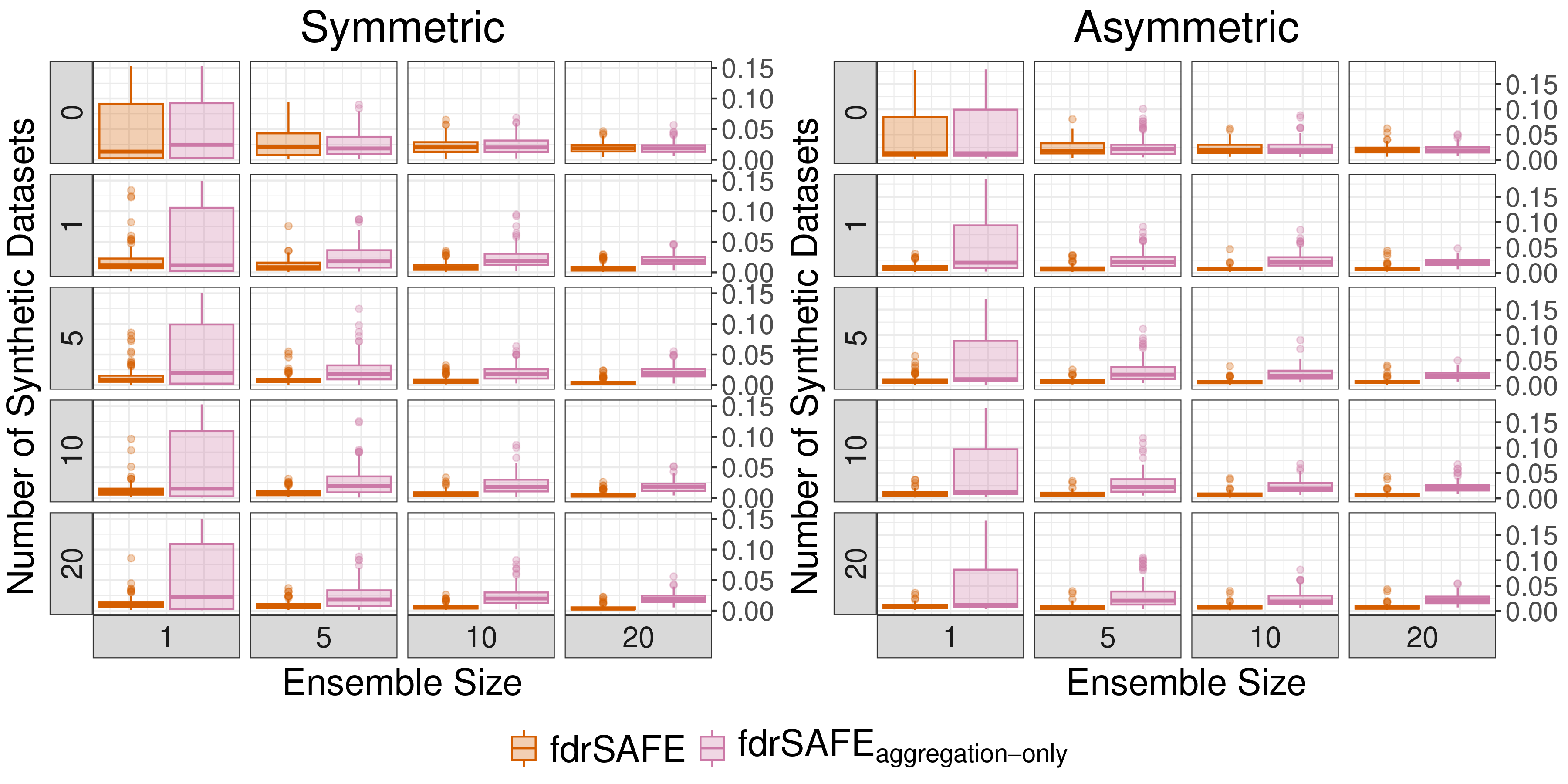}
    \caption{\textbf{Simulation results for varying ensemble size and number of synthetic datasets} comparing fdrSAFE to fdrSAFE$_{\text{aggregation-only}}$. Each box plot shows RMSE of fdr over 200 repetitions across the symmetric (left) and asymmetric (right) simulation studies.}
    \label{fig:increasing_n}
\end{figure}

Computational power and time requirements increase with the number of synthetic datasets, $N$, and model size increases with ensemble size, $m$. To see the effects of these values on performance and determine optimal values of $N$ and $m$, we compare our new methods fdrSAFE and its ablation fdrSAFE$_{\text{aggregation-only}}$ with varying number of synthetic datasets and ensemble sizes on the symmetric and asymmetric simulation studies (Figure \ref{fig:increasing_n}).

In both studies, fdrSAFE improves as ensemble size increases and as the number of synthetic datasets increases. fdrSAFE$_{\text{aggregation-only}}$ is unchanged by the number of synthetic datasets and performs worse as ensemble size increases (albeit with less variability). Around 10 synthetic datasets and an ensemble size of 10, the performance of fdrSAFE seems to plateau. We move forward with 10 synthetic datasets and an ensemble size of 10, though these can be user-specified with our R package, controlled with the {\tt n\_synthetic} and {\tt ensemble\_size} parameters.

\newpage
\section{Simulation Studies: Additional Results}

\subsection{Median Metrics}

Table \ref{tab:median_metrics} reports the median metrics for each method and study setups across 200 repetitions.

The results demonstrate fdrSAFE's robust near optimality, as it attains either the best of each metric, or very close to the best. In contrast, the baseline approaches performances vary depending on the simulation study.

\begin{table}[H]
    \centering
    \begin{tabular}{|l|l|r|r|r|r|r|}
    \hline
    \textbf{Study} & \textbf{Method} & \textbf{fdr/Fdr RMSE} & \textbf{Test Label} & \textbf{PR AUC} &\textbf{ ROC AUC} & \textbf{Estimated}\\
     &&& \textbf{Brier Score} && &\textbf{Prop. Null}\\

    \hline
    \textbf{Symmetric} & fdrtool & 0.230 & 0.109 & 0.915 & 0.806 & 0.919\\
     & locfdr & 0.200 & 0.097 & 0.941 & 0.871 & 0.916\\
     & qvalue & {\cellcolor[rgb]{0.753,0.753,0.753}} 0.046 & {\cellcolor[rgb]{0.753,0.753,0.753}}0.061 & {\cellcolor[rgb]{0.753,0.753,0.753}}0.992 & {\cellcolor[rgb]{0.753,0.753,0.753}}0.968 & {\cellcolor[rgb]{0.753,0.753,0.753}}0.797\\
     & fdrSAFE & 0.071 & 0.063 & {\cellcolor[rgb]{0.753,0.753,0.753}}0.992 & 0.966 & 0.843\\
     & fdrSAFE$_{\text{selection-only}}$ & 0.094 & 0.067 & 0.991 & 0.963 & 0.861\\
     & fdrSAFE$_{\text{aggregation-only}}$ & 0.144 & 0.078 & {\cellcolor[rgb]{0.753,0.753,0.753}}0.992 & {\cellcolor[rgb]{0.753,0.753,0.753}}0.968 & 0.873\\
     & fdrSAFE$_{\text{aggregation-all}}$ & 0.142 & 0.077 & {\cellcolor[rgb]{0.753,0.753,0.753}}0.992 & {\cellcolor[rgb]{0.753,0.753,0.753}}0.968 & 0.868\\
     & oracle ensemble & 0.031 & 0.059 & 0.992 & 0.968 & 0.794\\
     & oracle & 0.031 & 0.059 & 0.992 & 0.968 & 0.794\\
    \hline
    \textbf{Asymmetric} & fdrtool & 0.244 & 0.088 & 0.928 & 0.841 & 0.893\\
     & locfdr & 0.127 & 0.047 & 0.982 & 0.961 & 0.860\\
     & qvalue & 0.095 & 0.039 & {\cellcolor[rgb]{0.753,0.753,0.753}}0.997 & 0.986 & 0.803\\
     & fdrSAFE & {\cellcolor[rgb]{0.753,0.753,0.753}}0.086 & {\cellcolor[rgb]{0.753,0.753,0.753}}0.037 & {\cellcolor[rgb]{0.753,0.753,0.753}}0.997 & {\cellcolor[rgb]{0.753,0.753,0.753}}0.989 & 0.813\\
     & fdrSAFE$_{\text{selection-only}}$ & 0.096 & 0.038 & {\cellcolor[rgb]{0.753,0.753,0.753}}0.997 & 0.988 & {\cellcolor[rgb]{0.753,0.753,0.753}}0.799\\
     & fdrSAFE$_{\text{aggregation-only}}$ & 0.141 & 0.051 & {\cellcolor[rgb]{0.753,0.753,0.753}}0.997 & 0.988 & 0.851\\
     & fdrSAFE$_{\text{aggregation-all}}$ & 0.143 & 0.050 & {\cellcolor[rgb]{0.753,0.753,0.753}}0.997 & 0.988 & 0.854\\
     & oracle ensemble & 0.054 & 0.033 & 0.998 & 0.991 & 0.849\\
     & oracle & 0.051 & 0.032 & 0.998 & 0.991 & 0.883\\
    \hline
    \textbf{Curated Ovarian} & fdrtool & 0.174 & 0.091 & 0.920 & 0.824 & 0.891\\
    \textbf{Data-Based}& locfdr & 0.035 & 0.047 & 0.979 & 0.953 & 0.838\\
    & qvalue & 0.071 & 0.052 & 0.977 & 0.944 & {\cellcolor[rgb]{0.753,0.753,0.753}}0.805\\
    & fdrSAFE & {\cellcolor[rgb]{0.753,0.753,0.753}}0.029 & {\cellcolor[rgb]{0.753,0.753,0.753}}0.046 & 0.984 & 0.959 & 0.834\\
     & fdrSAFE$_{\text{selection-only}}$ & 0.033 & 0.048 & 0.979 & 0.952 & 0.830\\
     & fdrSAFE$_{\text{aggregation-only}}$ & 0.149 & 0.060 & 0.984 & 0.960 & 0.854\\
     & fdrSAFE$_{\text{aggregation-all}}$ & 0.152 & 0.059 & {\cellcolor[rgb]{0.753,0.753,0.753}}0.985 & {\cellcolor[rgb]{0.753,0.753,0.753}}0.961 & 0.851\\
     & oracle ensemble & 0.015 & 0.046 & 0.983 & 0.957 & 0.830\\
     & oracle & 0.012 & 0.046 & 0.981 & 0.955 & 0.826\\
    \hline
    \end{tabular}
    \caption{\textbf{Simulation study results} showing median metrics of fdrSAFE and each of its ablations fdrSAFE$_{\text{selection-only}}$, fdrSAFE$_{\text{aggregation-only}}$, and fdrSAFE$_{\text{aggregation-all}}$ versus each of the baselines locfdr, fdrtool, and qvalue. The best non-oracle value of each metric in each study is highlighted (closest to 0.8 for estimated proportion null). Companion to Figure 1 recording median values from each boxplot.}
    \label{tab:median_metrics}
\end{table}

\newpage
\section{Data Application: Additional Results}

\subsection{Classification Results} \label{sec:supp_classif}

In addition to the metrics based on nominal fdr values, we can examine the downstream utility of $\widehat{\mathbf{fdr}}(\mathbf u)$ in classifying hypotheses as null or alternative. The choice of cutoff for classification is very subjective, so we report results under four cutoff options: first, an ``oracle'' cutoff chosen so the true $\pi_0$ proportion of tests are classified as null, second, a cutoff chosen so that estimated $\hat\pi_0$ proportion of tests are classified as null, third, the standard cutoff of 0.2 recommended in \citet{efron2007}, and fourth, a cutoff chosen so that the estimated global FDR is fixed at 0.05. In summary, cutoffs are:
\begin{equation}
\begin{aligned}
\label{eq:cutoff}
    c^{(\text{oracle }{\pi}_0)}(\widehat{\mathbf{fdr}}(\mathbf u)) & = (1- \pi_0)*100\text{th quantile of }\widehat{\mathbf{fdr}}(\mathbf u)\\
    c^{(\widehat{\pi}_0(\mathbf u))}(\widehat{\mathbf{fdr}}(\mathbf u)) & = (1-\widehat \pi_0(\mathbf u))*100\text{th quantile of }\widehat{\mathbf{fdr}}\\
    c^{(standard)} &= 0.2\\
    c^{(\widehat{\text{FDR}}(\mathbf u) = 0.05)}(\widehat{\mathbf{fdr}}(\mathbf u)) & = \min_{c} \left\{ c: \frac{\sum_{i = 1}^I \widehat{\text{fdr}}_i(\mathbf u) \cdot \mathbf 1(\widehat{\text{fdr}}_i(\mathbf u) \le c)}{\sum_{i = 1}^I \mathbf 1(\widehat{\text{fdr}}_i(\mathbf u) \le c)} \le 0.05 \right\}
\end{aligned}
\end{equation}
and classification rules, for a given cutoff $c$, are:
\begin{equation}
 \hat{l}_i = \begin{cases}
    \hbox{null} & \hbox{ if } \widehat{\hbox{fdr}}_i(\mathbf u) > c\\
    \hbox{not null} & \hbox{ if } \widehat{\hbox{fdr}}_i(\mathbf u) \le c.
    \end{cases}
\end{equation}

Table \ref{tab:data_classif} shows classification results in terms of global FDR, the number of true and false discoveries, and sensitivity---or the proportion of all true positives that are discovered.

\begin{table}[H]
    \begin{center}
    \begin{tabular}{|r|l|l|l|l|} 
        \cline{2-5}
            \multicolumn{1}{l|}{} & \textbf{fdrSAFE} & \textbf{locfdr} & \textbf{fdrtool} & \textbf{qvalue} \\
        \hline
            \multicolumn{1}{|l|}{\textbf{$\pi_0$-Based Cutoff}} & & & & \\
            cutoff $c^{(\pi_0)}(\widehat{\mathbf{fdr}}(\mathbf u))$ & 0.44 & 0.13 & 1 & 1 \\
            FDR & 0.147 & 0.150 & 0.638 & 0.638 \\
            Sensitivity & 0.853 & 0.850 & 0.999 & 1 \\
            True Discoveries & 1658 & 1652 & 1943 & 1944 \\
            False Discoveries & 286 & 292 & 3426 & 3426\\ 
        \hline
            \multicolumn{1}{|l|}{\textbf{$\hat{\pi}_0(\mathbf u)$-Based Cutoff}} & & & & \\
            cutoff $c^{(\hat\pi_0(\mathbf u))}(\widehat{\mathbf{fdr}}(\mathbf u))$ & 0.13 & 0.23 & 0.12 & 0 \\
            FDR & 0.097 & 0.197 & 0.096 & 0 \\
            Sensitivity & 0.721 & 0.908 & 0.705 & 0.001 \\
            True Discoveries & 1401 & 1765 & 1370 & 2 \\
            False Discoveries & 150 & 434 & 146 & 0 \\ 
        \hline
            \multicolumn{1}{|l|}{\textbf{Standard 0.2 Cutoff}} & & & & \\
            cutoff $c^{(\text{standard})}$ & 0.2 & 0.2 & 0.2 & 0.2 \\
            FDR & 0.111 & 0.176 & 0.100 & 0.082 \\
            Sensitivity & 0.759 & 0.899 & 0.708 & 0.638 \\
            True Discoveries & 1475 & 1747 & 1377 & 1240\\
            False Discoveries & 184 & 372 & 153 & 110 \\
        \hline
            \multicolumn{1}{|l|}{$\widehat{\text{FDR}}(\mathbf u)=0.05$\textbf{ Cutoff}} & & & & \\
            cutoff $c^{(\widehat{\text{FDR}}(\mathbf u)=0.05)}(\widehat{\mathbf{fdr}}(\mathbf u))$ & 0.37 & 0.36 & 1 & 0.17 \\
            FDR & 0.138 & 0.241 & 0.117 & 0.078 \\
            Sensitivity & 0.817 & 0.928 & 0.771 & 0.598 \\
            True Discoveries & 1588 & 1804 & 1498 & 1163\\
            False Discoveries & 255 & 574 & 189 & 99\\ 
        \hline
    \end{tabular}

    \end{center}
    \caption{\textbf{Classification results on experimental application} comparing fdrSAFE to each of the baseline methods.
    $\hat \pi_0$ should be compared to the true value $\pi_0 = 0.638$. Classification metrics show the downstream utility of estimated fdr values in classifying tests as null (small fdr) or non-null (large fdr). FDR is global false discovery rate, the proportion of discovered tests that are actually null. Sensitivity is the proportion of non-null tests that were correctly discovered.}
    \label{tab:data_classif}
\end{table}

\subsection{Subset Analysis with Varying $\pi_0$} \label{sec:subset}

The Platinum Spike dataset allows us to look at real differential gene expression data and to perform multiple hypothesis testing with known labels of which hypotheses are null and alternative. To evaluate fdrSAFE and benchmark methods on real differential expression data with a range of $\pi_0$ values, we sample subsets of genes to construct datasets with varying $\pi_0$. We consider true $\pi_0$ values between 0.6 and 0.95. For each $\pi_0$, we sample ten subsets of 1000 genes.

\begin{figure}[H]
    \centering
    \includegraphics[width = 0.8\linewidth]{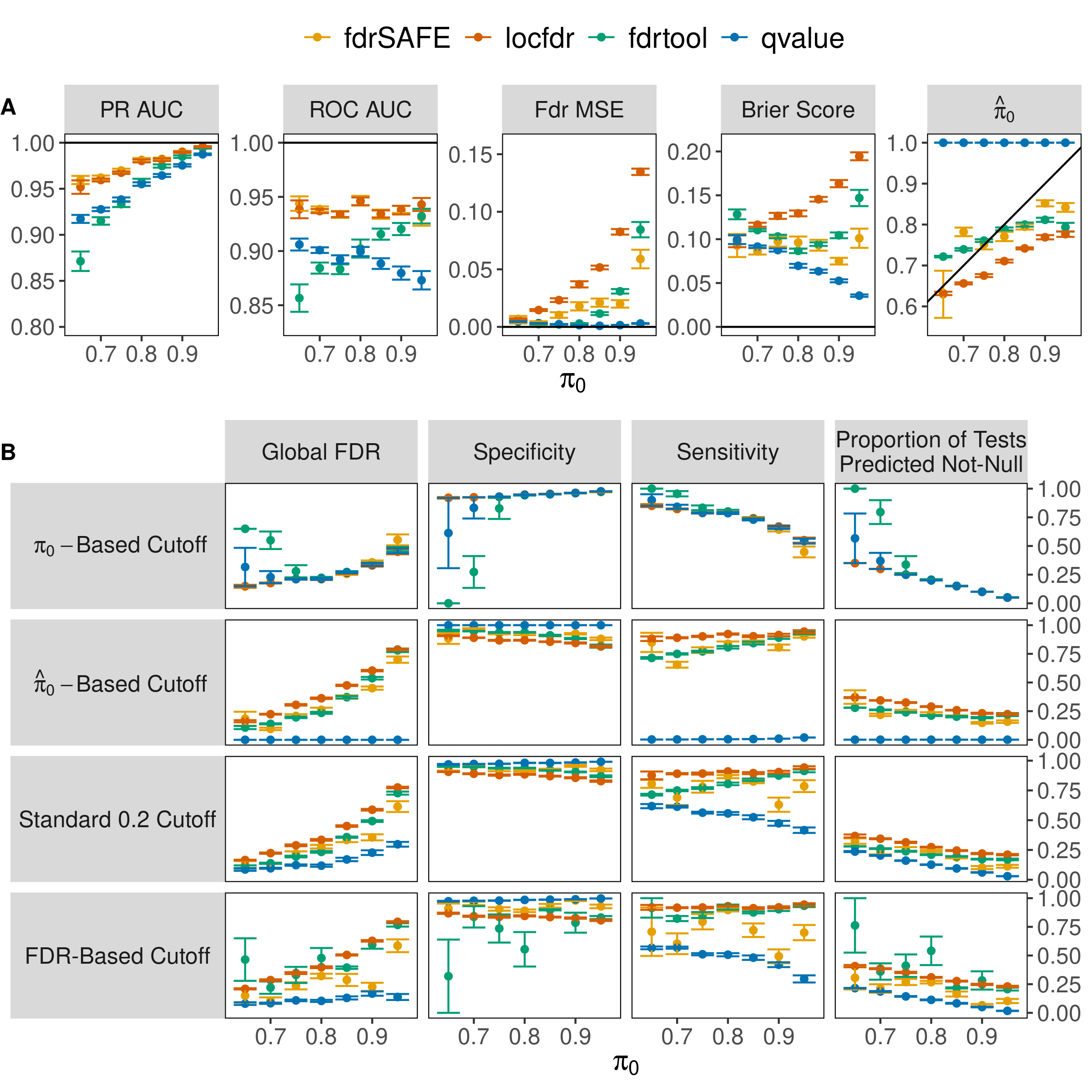}
    \caption{\textbf{Subset results on Platinum Spike dataset.} For each $\pi_0$, 10 subsets of the Platinum spike dataset were generated for analysis. Mean and standard error of each quantity is plotted for every $\pi_0$.  \textbf{A)} Metrics on $\widehat{\mathbf{fdr}}(\mathbf u)$. \textbf{B)} Classification metrics with four cutoff strategies (see Appendix Section \ref{sec:supp_classif}). We note that when $\pi_0$ is low, the poor results of fdrtool and qvalue for the $\pi_0$-based cutoff, as well as of fdrtool for the FDR-based cutoff, are due to a large proportion of their $\widehat{\mathbf{fdr}}(\mathbf u)$ values at exactly 1. The cutoff is then also 1, and all hypotheses are classified as non-null.}
    \label{fig:sens}
\end{figure}

For metrics on $\widehat{\mathbf{fdr}}(\mathbf u)$, fdrSAFE performs very well across values of $\pi_0$ (Figure \ref{fig:sens}A). Importantly, for Fdr MSE, PR AUC, and ROC AUC, we see that fdrSAFE is consistent in its high performance even when other methods' performances vary with $\pi_0$.

Classification results using a cutoff based on the true $\pi_0$ show that for all methods and most values of $\pi_0$, $\widehat{\mathbf{fdr}}(\mathbf u)$ can lead to good classification metrics given the right cutoff value (Figure \ref{fig:sens}B). However, in practice, we don't have access to the true $\pi_0$ values. The classification results using cutoffs based on the estimated $\widehat{\pi}_0(\mathbf u)$ show that in practice, fdrSAFE is well equipped to make an accurate cutoff decision because of its accurate $\widehat{\pi}_0(\mathbf u)$ estimates. fdrSAFE has the best global FDR and specificity without predicting all hypotheses as null like qvalue. With the fixed 0.2 and FDR-based cutoffs, qvalue has lower global FDR, but at the cost of reduced sensitivity.

\newpage
\subsection{Data Distribution and qvalue's Estimation of $\pi_0$}\label{sec:qvalue_distribution}

\begin{figure}[H]
    \centering
    \includegraphics[width=0.8\linewidth]{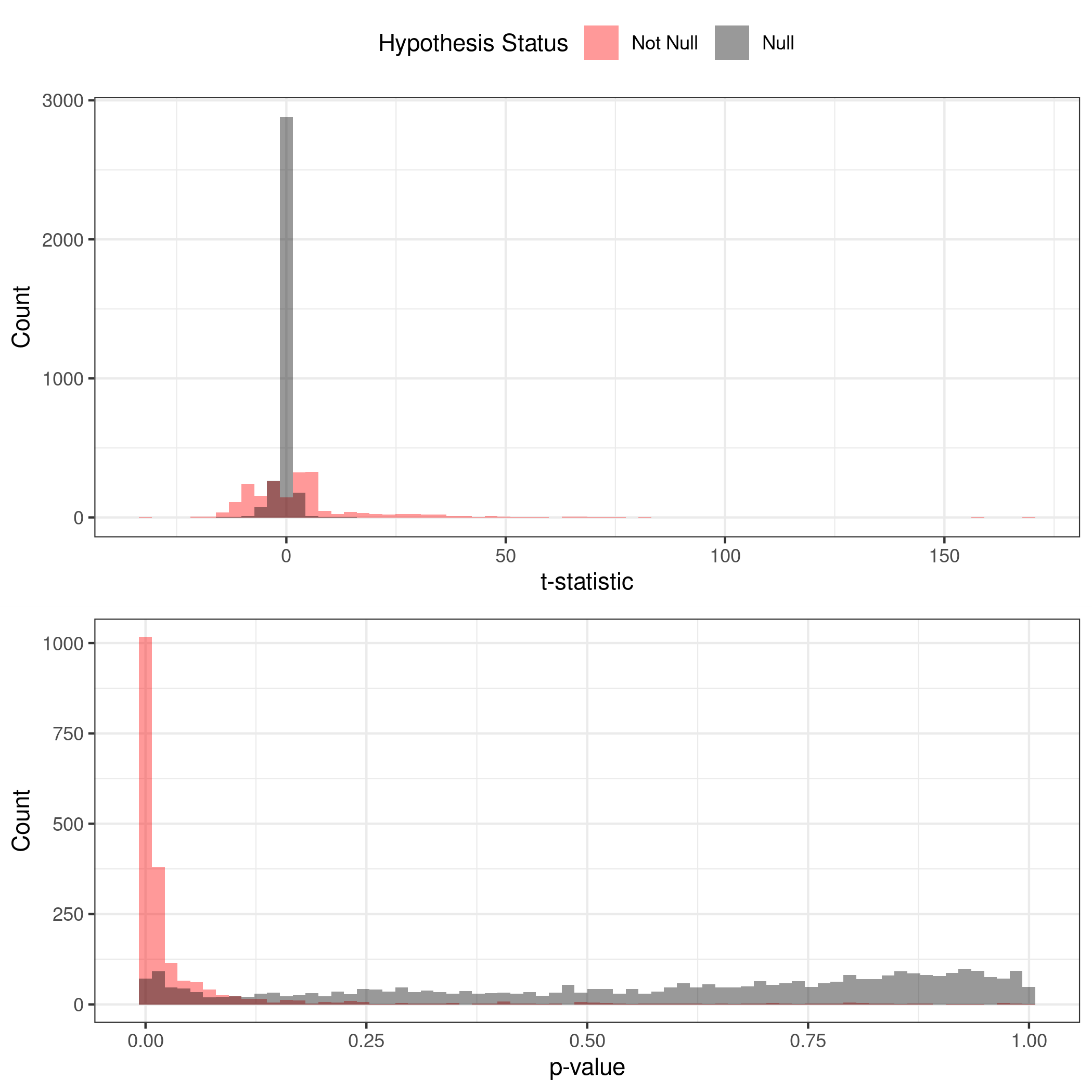}
    \caption{\textbf{Platinum Spike dataset distribution} of t-statistics and p-values, separated by hypothesis status.}
    \label{fig:data_dist}
\end{figure}

On the Platinum Spike dataset, qvalue consistently estimates proportion null at 1, even in the subset analysis of Section \ref{sec:subset}. This behavior is explained by the bulk of null p-values near 1 in Figure \ref{fig:data_dist} and the particular way qvalue computes $\hat\pi_0$, described next.

qvalue assumes that null p-values follow a uniform distribution and uses the following equality for a conservative estimate of the proportion null:
\begin{align*}
    \hat \pi_0(\lambda) = \frac{\sum_{i = 1}^I \mathbf 1(p_i > \lambda)}{I\cdot (1-\lambda)} = \frac{\text{\# classified null with cutoff }\lambda}{\text{\# expected to be greater than }\lambda \text{ if all tests are null}}
\end{align*}
where $p_i$ is the p-value computed from test statistic $u_i$. $\hat \pi_0(\lambda)$ estimates $\pi_0$ correctly if we are confident that all tests with $p_i > \lambda$ are null. In this sense, $\hat \pi_0(1)$ should approximate the overall $\pi_0$.

With qvalue, $\hat \pi_0(\lambda)$ is computed for increasing values of $\lambda$, a natural cubic spline is fit to $\hat \pi_0(\lambda)$ with respect to $\lambda$, and this spline is evaluated at $\lambda = 1$ for the final estimate of $\hat \pi_0$. However, in cases like ours with a bulk of large null p-values breaking uniformity, $\hat \pi_0(\lambda)$ can exceed 1, and qvalue caps the final $\hat \pi_0$ estimate at 1.

\end{document}